\documentclass{emulateapj}
\usepackage{psfig}

\def\la{\mathrel{\mathpalette\fun <}}

\def\fun#1#2{\lower3.6pt\vbox{\baselineskip0pt\lineskip.9pt
        \ialign{$\mathsurround=0pt#1\hfill##\hfil$\crcr#2\crcr\sim\crcr}}}
\def\be#1{\begin{equation}\label{eq:#1}}
\def\ee{\end{equation}}

\def\bea#1{\begin{eqnarray}\label{eq:#1}}
\def\ee{\end{equation}}
\def\eea{\end{eqnarray}}
\def\la{\mathrel{\mathpalette\fun <}}

\def\kmskpc{{\rm ~km~s^{-1}~kpc^{-1}}}
\def\msol{{\rm ~M_\odot}}
\def\kms{{\rm km~s^{-1}}}

\begin{document}
\title{Warps and Bars from the External Tidal Torques of Tumbling Dark Halos}

\author{
John Dubinski\altaffilmark{1} and
Dalia Chakrabarty\altaffilmark{2} and
}

\altaffiltext{1}{Department of Astronomy and Astrophysics,
University of Toronto, 50 St. George Street, Toronto, ON M5S 3H4, Canada;
E-mail: dubinski@astro.utoronto.ca}
\altaffiltext{2}{
School of Physics $\&$ Astronomy, University of Nottingham, Nottingham NG7
2RD;
E-mail: dalia.chakrabarty@nottingham.ac.uk}



\label{firstpage}

\begin{abstract}
The dark matter halos in $\Lambda$CDM cosmological simulations are
triaxial and highly flattened.  In many cases, these triaxial
equilibria are also tumbling slowly, typically about their short axes,
with periods of order a Hubble time.  Halos may therefore exert a
slowly changing external torque on spiral galaxies that can affect
their dynamical evolution in interesting ways.  We examine the effect
of the external torques exerted by a tumbling quadrupolar tidal field
on the evolution of spiral galaxies using N-body simulations with
realistic, disk galaxy models.  We measure the amplitude of the
external quadrupole moments of dark halos in cosmological simulations
and use these to force disk galaxy models in a series of N-body
experiments for a range of pattern speeds.  We find that the torques
are strong enough to induce long lived transient warps in disks
similar to those observed in real spirals and also induce the bar
instability at later times in some galaxy models that are otherwise
stable for long periods of time in isolation.  We also observe forced
spiral structure near the edge of the disk where normally self-gravity
is too weak to be responsible for such structure.  This overlooked
influence of dark halos may well be responsible for many of the
peculiar aspects of disk galaxy dynamics.
\end{abstract}


\maketitle

\section{Introduction}
The nearly flat rotation curves of spiral galaxies are direct evidence
for the existence of massive dark halos around the galaxies, within
the paradigm of Newtonian-Einsteinian gravity \citep{rob75, rubin_79, rubin_82,
rubin_85, van85, salucci_00, sofue_rubin}. The dark halos forming in cold dark
matter cosmological models have been identified with the inferred dark
halos of galaxies; a substantial amount of work has gone into trying
to test for the consistency between reality and simulation
\citep[e.g.,][]{barnes_87, frenk_88, dubinski_91, nav96, bul01, neto07}.
Most work has focused on the spherically-averaged density
profile and the implications of this profile for galactic rotation
curves. Although debate still continues on the value of the slope of
the inner cusp \citep{gentile_04, ferreras_07} and whether or not the
inferred rotation curves are consistent with real ones
\citep{moo94,mcgaugh_98, mcgaugh_03}, it seems that the consistency between
the real and simulated dark halos is quite strong and that the cold
dark matter (CDM) paradigm is in reasonable shape in this sense
\citep{primack_07}.

Other work has focused on secondary dynamical effects of dark halos
that arise from their triaxial nature and their implication for other
properties in disks, namely, oval distortions, warps and perhaps bars
\citep{bin78,sparke_84,fra_92,sackett_94,kuijken_94,weinberg_98,
sellwood_03,shlosman_06}.  The flattened potential of triaxial halos has
been recognized in some work to induce oval distortions to otherwise
circular disks, as well as vary the velocity along the orbit
\citep{hayashi_07}. The orbits of stars in the disk become flattened
in a direction orthogonal to the halo density. However, the detection
of such oval distortions and non-circular velocities is marginal at
best, suggesting that halos may be nearly axisymmetric in the plane of
an ordinary, high surface-brightness spiral galaxy. On the other hand,
the effects of non-circularity in dwarfs and low-surface brightness
galaxies may be stronger \citep{valenzuela_07, hayashi_06}.

Another source of complex dynamical effects is the probable
misalignment between a disk and a triaxial dark halo. Disk-halo
misalignment has been argued to be the origin for warps seen in most
edge-on spiral galaxies. A misalignment implies that the disk will
feel a torque from the halo and the nearly circular orbits will
therefore precess. If the disk has self-gravity and the halo potential
is static, warped modes can arise \citep[e.g.,][]{toomre_warps,
sparke_88, kui91}. A fatal flaw in this hypothesis is that
gravitational interactions between the disk and the halo are strong,
with the precessing disk experiencing dynamical friction from the halo
\citep[e.g.,][]{nelson_95, binney_98}. One therefore expects the
disk and halo to become aligned with one another within a few
dynamical times and this is indeed borne out in experiments
\citep{dubinski_kuijken}. Thus it has been suggested that it is
the outer halo that is misaligned with the disk, while a tight
coupling exists between the inner halo and the disk \citep{bailin_05, 
binney_warps_07}.

The origin of galactic warps then remains a perplexing issue but
alternative scenarios suggest a transient origin.
Other ideas suggest that the cosmic infall of gas and dark matter
alters the relative orientation the disk and dark halo due 
to addition of angular momentum \citep[e.g.,][]{ostriker89,deb99,binney_99} 
and the resulting torques on the disk lead to transient warping.
\citet{vanderkruit_07} also considers the possibility of the onset of a
late warped outer disk, due to gas infall, in a configuration that is
independent of the relatively younger inner truncated stellar disk.
The infall picture has also been recently used by \citet{juntai_06} to
reproduce the warping of a simulated disk galaxy; in fact, their
simulations corroborate the result that the line-of-node of the warp
forms a leading spiral. 

Tidal fields from satellite galaxies are known to excite dramatic
behavior in spirals; \cite{m31} invoke the possibility of a warp or
overdensity induced by satellite interaction to explain the origin of
the secondary cold population that they identify in M31 while the
grand design spiral in M51 is inferred to be the result of it's
interaction with its companion NGC~5194 \citep{too72,salo_00}. 
The Magellanic origin of the Galactic warp \citep{weinberg_98,garcia-ruiz}
is revisited by \citet{weinberg_06} in which they suggest that the
warp is formed as a joint handiwork of the tidal field of the
Magellanic Clouds as well as the effect of the distortions that such
has on the Galactic halo.  The overall picture that emerges from these
studies is that the source of the warping torque lies with asymmetries
in the halo, at radii beyond the gravitational influence of the disk.

Dark halos formed in simulations performed within the CDM scenario
suggest that not only do they often settle into triaxial figures of
equilibrium but that they can also tumble, much like a rigid body
\citep{dubinski_92, pfitzner_99, bai04}. Static and
tumbling equilibrium are both perfectly valid in analogy to the two
solutions for the homogeneous Jacobi ellipsoids \citep{chandra_69}
which include a tumbling and static solution with internal
circulation.  \citet{bai04} find a log normal distribution
of halo pattern speeds with a mean value of about $\Omega_p=0.15 \kmskpc$, 
corresponding to a tumbling period of 44 Gyr. A typical halo will then 
tumble through 120 degrees over a Hubble time.
The tumbling periods of dark halos are rather long
but nevertheless, a picture emerges in which disk galaxies are
embedded within slowly tumbling, triaxial halos with rotation axes
that are probably slightly misaligned with the disk rotation axes. It
is likely that the inner halo and disk are aligned with each other but
the outer halo (say beyond 100 kpc) is tumbling slowly and thereby
affecting the disk with its tidal field. A pertinent question to ask
then is how strong are the torques that a tumbling dark halo exerts
on the disk at the center and whether or not these torques can have a
significant influence on the evolution of a spiral galaxy over a Hubble
time?

In this paper, we carry out a series of experiments to study the
effect of a slowly tumbling, external quadrupolar potential, on the
evolution of a disk, with the primary goal of inducing galactic warps.  
We first measure the expected strength of the quadrupolar tidal field 
directly from dark matter halos within $\Lambda$CDM cosmological simulations 
and subsequently estimate the torques expected on exponential disks.  
We find that the typical torques 
from cosmological dark halos are generally an order of magnitude larger 
than the effect of the LMC, rendering them more effective in trigerring 
something dynamically interesting. 

Guided by such experimentally motivated numbers, we set up simulations
of ideal galactic models embedded in nearly spherical halos,
forced slowly by an external quadrupolar tidal field. We discover in
these experiments that it is fairly simple to make warps of amplitudes
similar to the observed ones.

We also find that the gradual tidal forcing of tumbling dark halos has
the effect of pushing an otherwise stable disk into a region where it
becomes subject to the bar instability at late times.  This is a new
bar triggering mechanism and may be an additional factor that affects
the evolution of the fraction of barred galaxies over cosmic history
\citep[e.g.,][]{curir08,she08}.

The plan of the paper is as follows. In \S 2, we calculate the
amplitude of the torque on a galactic disk, expected from the external
tidal field of the triaxial dark halo, as extracted from cosmological
$N$-body simulations. In \S 3, we present
$N$-body models of M31 and the Milky-Way that are forced by 
the external quadrupolar field with the same strength expected from
cosmological simulations.
In \S 4, we present the results of these simulations while in \S 5,
present analytical models of rigid disks forced by external tidal
fields, to quantify these effects. In \S 6 we conclude with a
discussion of the implications of these results for warps, bars, and
spirals in real galaxies.

\section{Tidal Torques on Disks from Cosmological Dark Halos}

The dark halos in CDM cosmological models are highly flattened and
triaxial with typical axis ratios $c/a=0.4$ and
$b/a=0.6$ \citep[e.g.,][]{barnes_87, frenk_88, dubinski_91, warren_92, jin02}.
Analysis of the halo spin parameter shows a tendency for alignment
with the minor axis, suggesting that a disk that forms within a dark
halo is likely to have its own spin angular momentum closely aligned
with the halo minor axis \citep[e.g.,][]{dubinski_92,warren_92,
dubinski_kuijken, binney_warps_07}. The angular momentum
within halos is distributed between internal streaming and tumbling
motion.  Early studies showed that halos are slowly tumbling through
space with long periods. More recent quantitative analyses in standard
$\Lambda$CDM cosmological models show that the distribution of
tumbling frequencies is roughly log-normal \citep{bai04}
and peaked at about $0.15 \kmskpc$ with $\sigma=0.83$.
The dissipative infall of the baryonic matter during galaxy formation
can also modify the halo, both increasing its central concentration 
\citep[e.g.,][]{blu84} as well as rounding out the shape by increasing $b/a$
within the region of influence of the forming disk
while leaving $c/a$ about the same \citep{dub94,kaz04}.  The axis ratios
of the volumetric density contours at radii beyond the influence of the
disk are less affected and remain about the same as the initial values.

Disks and halos are also found to be tightly coupled through dynamical
friction in the inner regions of galaxies (Dubinski \& Kuijken 1995)
so one would expect the disk to lie in the principal plane of the
halo, probably with the spin vectors of the disk and halo aligned
\citep{bailin_05, libeskind_07}. However, misalignment may persist at
larger radii and the slow tumbling of a halo could lead to an external
torque on the disk.  Although current self-consistent simulations 
of galaxy formation have not fully addressed this point, we make 
the hypothesis that the disks are aligned with the principal planes
of their inner halos through dynamical friction while the outer 
halos remain misaligned and may slowly be changing orientation
with the tumbling frequencies measured in cosmological
dark matter simulations.

We estimate this torque through the analysis of more than 2000 dark
halos extracted from a cosmological dark matter simulation with
near standard parameters 
$\Omega_\Lambda=0.7$, $\Omega_m=0.3$, $\sigma_8=0.9$ and $h=0.7$
\citep{wmap3}. The simulation contains $512^3$ particles in a cube of
dimension $L=70h^{-1}$Mpc and was run using the GOTPM cosmological N-body
code \citep{dub04}. The
simulation was run from $z=70$ using 2800 equal time steps. The
co-moving particle softening length was set to 3.5$h^{-1}$kpc.  Halos
are extracted in a two stage process; we use a friends-of-friends
algorithm to determine the dense centers of candidate halos.  These
regions could either be associated with an isolated halo or merging
pair or group.  The particles in a spherical region surrounding these
dense centers are extracted out to roughly the distance where the
density drops to 200 times the critical density and then
analyzed separately.  We fit Navarro, Frenk \& White 
(1996, NFW) profiles
to all these candidates but
retain only those which have fit profiles that lie within a threshold
$\chi^2$.  In this way, the original sample of about 2700 halos is
reduced to a sample of about 2200 halos that have well-defined
profiles, undisturbed by major mergers or significant asymmetries.  We
then take a careful look at this sample to learn something about halo
shapes and torques.

\begin{figure}
\plotone{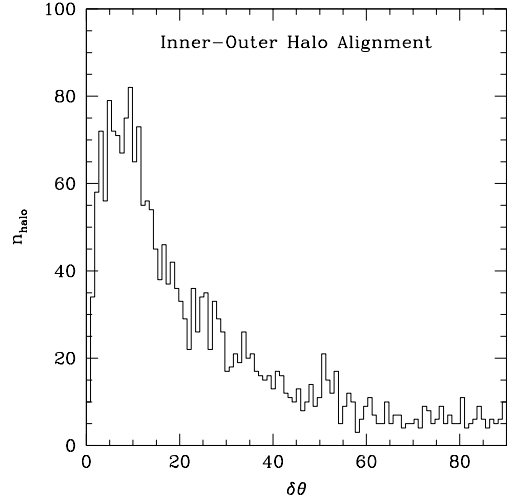}
\caption{{The distribution of the relative alignment between the 
inner and outer halo from the sample of cosmological dark halos.
We measured the normalized moments of inertia for particles in the
inner halo with $r < r_s$ and the outer halo for particles
in the range $r_s < r < 2 r_s$.  Diagonalization of the inertia
tensor determines the direction of the principle axes.  We calculate
the angle between the minor axes of the inner and outer halos $\delta \theta$
and plot the distribution.  As expected the inner and outer halos are closely
aligned though the median of distribution around 20$^\circ$ implies a
significant misalignment between in the inner and outer halos for at
least half of the dark halos.}
\label{fig-halo-align}
}
\end{figure}

Many studies have determined the distribution of ellipsoidal
shape of halos \citep{frenk_88, dubinski_91, warren_92, cole_lacey,
jin02,bailin_steinmetz}.
We take this analysis one step further and
calculate the distribution in expected strengths of tidal torques
on a disk that may be misaligned with the outer halo. An explicit
assumption is that the disk and halo are aligned and will remain so because
of dynamical friction in the inner regions while misalignments between the
outer halo will persist and change dynamically as the halo tumbles through
space.  We find that the relative orientations of the inner ($r<r_s$)
and outer halo ($r>r_s$) is greater than 20$^\circ$ for half of the 
halos in our cosmological simulation (Fig.~\ref{fig-halo-align}).
We therefore expect significant misalignment that can lead to
torques on the disk from the outer halo.
We proceed by
analysing the potential of the outer halo using a standard multipole
expansion method.

We are interested in the possible torques at the centre of a disk
tilted with respect to the outer halo. Previous work has shown that
disks and inner halos are aligned within a radius roughly equal to
the extent of the disk, though such alignment may extend further
\citep{dubinski_kuijken,binney_98}. Any
torque acting on the disk from the halo probably then arises from
misalignments beyond this radius.  It is useful to quantify these
effects more concretely.
According to the NFW model, 
dark matter halos extend to the virial radius $r_{200}=c r_s$ where
$c$ is the concentration, $r_s$ is the scale radius of the density profile
and the mean enclosed density is 200 times the critical density $\rho_c=3
H_0^2/8\pi G$.  At $r_{200}$, halos have characteristic mass $M_{200}$ and
circular velocity $v_{200}^2 = G M_{200}/r_{200}$.  Typical
concentrations for galactic scale halos are $c \approx 15$
\citep[e.g.,][]{bul01} though they can vary considerably with a scatter
$\Delta(\log c) = 0.18$.  A galaxy halo is usually characterized by the 
peak value of the rotation curve $v_{c,max}$ rather than $v_{200}$.
Analysis of the NFW
rotation curve shows that the maximum value of the rotation curve
$v_{c,max}$ occurs at
$r=2.16 r_s$ and is related to $v_{200}$ :  
\begin{equation}
v_{200} = 2.15~v_{c,max}~f(c)
\label{eq-vcmax}
\end{equation}
with
\begin{equation}
f(c) = c^{-1/2}\left ( \ln(1+c) - \frac{c}{1+c} \right)^{1/2}.
\end{equation}
Given the parameters $c$ and $v_{c,max}$ we find $v_{200}$ and 
then can determine
$r_{200}$ (or $r_s$) and $M_{200}$ through the usual identities
\citep{nav96}:
\begin{equation}
r_{200} = 100 \left ( \frac {v_{200}}{100~\kms} \right ) h^{-1}~{\rm kpc}
\end{equation}
\begin{equation}
M_{200} = 2.325\times 10^{11} \left ( \frac{r_{200}}{100~{\rm kpc}} \right) \left
( \frac{v_{200}}{100~\kms} \right )^2~\msol
\end{equation}

As a specific example, consider a Milky Way
sized NFW dark halo with with a $v_{c,max}=220 \kms$ and concentrations
with the range $c=10-20$.  Assuming $h=0.7$, 
these models will have virial masses in
the range $M_{200}=1.1-2.0 \times 10^{12} \msol$ and 
scale radii from $r_s=11-26$~kpc.
The scale radius $r_s$ is roughly the size of the disk and since
the disk and halo are tightly coupled within this region we expect
alignment.
We can expect misalignment and tidal torques from the halo mass 
distribution for halo mass beyond the edge of the disk
and so we need to calculate the
external component of the potential beyond $r > r_s$ to estimate the
tidal torques.

For a halo particle distribution, the external potential from matter
with $r>r_0$ at a point
$(r,\theta,\phi)$ with $r<r_0$ is given by the expression 
(e.g., Binney and Tremaine 1987),
\begin{equation}
\Phi_{ext}(r; r<r_0) = \displaystyle{
                                     -4\pi G\sum_{l=0}^{l=\infty} 
                                            \sum_{m=-l}^{m=l} c_{lm} r^l 
                                                              Y_l^m(\theta,\phi)
                                    },
\end{equation}
where $Y_l^m$ is the usual spherical harmonic function and the coefficients 
$c_{lm}$ are evaluated from the matter beyond $r > r_0$ through:
\begin{equation}
c_{lm}=\displaystyle{
                     \sum_{\alpha=1}^N  
                     \frac{m_\alpha r_\alpha^{-(l+1)}}{4\pi(2l+1)}
                     Y_l^{m*}(\theta_\alpha,\phi_\alpha)
                    }, \quad r_\alpha > r_0
\label{eqn:coeffs}
\end{equation}
with $\alpha$ indexing a list of the $N$ particles
that have $r>r_0$.  

For a dark halo, it is natural to compute these coefficients for the particles
with $r_0 = r_s$ and $2r_s$, as a reasonable measure of the external
torquing potential of a dark halo.

Taking the lead from CMB analysis, a useful way of quantifying the
strength of the halo tidal potential is through the parameter $c_l$
defined as:
\begin{equation}
c_l = \left( \sum_{m=-l}^{m=l} \frac{|c_{lm}|^2}{2l+1} \right)^{1/2}
\end{equation}
The monopole term will lead to zero torque and the dipole terms are
probably unimportant since dark halos tend to be ellipsoidal and do
not show a significant lopsided mode.  The most important terms for
torquing are the quadrupole terms $l=2$; one expects successive
even $l$ terms to have some effect but with increasingly lower
strengths. We therefore focus on the quadrupole terms that have a
strength given by the parameter $c_2(r>r_0)$ as the main source of
torque on a misaligned disk. We now calculate this value for our
sample of halos and determine its importance for disk torques.

Since the value of $c_2$ depends on the mass of the halo, we can
rescale all halos to some reference mass. We are interested mainly in
the distribution of the relative strength of the external tidal
potential in typical spiral galaxies; so we first renormalize the
halos to have the same scale radius and scale mass. The following
formalism is then undertaken with these halos.
\begin{enumerate}
\item We first fit the scale parameters $r_s$ and $M_s$ to the NFW density
profile in $\log \rho - \log r$ space with the potential-density pair
of the spherical NFW profile given by:
\begin{eqnarray}
\Phi(r) &=& -\frac{GM_s}{r} \log(1+r/r_s) \\
\rho(r) &=& \frac{M_s}{4\pi}\frac{1}{r(r+r_s)^2}
\end{eqnarray}
\item Next we determine $r_{200}$ as the radius containing the mean density
$200\rho_c$ and find $c=r_{200}/r_s$ and so compute $v_{c,max}$, $v_{200}$
and $M_{200}\equiv M_s [\ln(1+c) - c/(1+c)]$ from the potential.
\item
We rescale all halos to a putative Milky Way model
for our determination and call the rescaled tidal parameter
$c_{2,MW}$.
We use the parameters
$v_{c,max} = 220~\kms$, a concentration of $c=10$ which imply
$r_s=26$~kpc and $M_{200} = 2\times 10^{12}~\msol$ from the equations
above.
\end{enumerate}

To correlate this parameter to a specific example, we compute the
expected value of the quadrupole tidal parameter for a satellite of
the type of the Large Magellanic Cloud, (mass $M=10^{10}$ M$_\odot$ at
a distance of $R=50$ kpc). We refer to this tidal parameter as
$c_{2,LMC}$.  The LMC has a minor tidal
effect on the dynamics of the Galaxy at its current distance \citep[e.g.,.][]{hunter_toomre,
garcia-ruiz} but simulations of disk galaxy satellite encounters
indicate that interactions in which the orbit of the LMC intersects
the disk during a close encounter in the past may excite a strong tidally 
generated spiral \citep{weinberg_06}. 

A close encounter with the LMC that brings it to half its current distance 
is expected to correspond to a quadrupolar tidal field that is about 
8 times stronger since $c \sim r^{-3}$.
Thus, the ratio $q_{tidal}=c_{2,MW}/c_{2,LMC}$ is a good indicator
of the active strength of the halo tidal field on Galactic dynamical
evolution. If $q_{tidal}\approx 1$ then the tidal field is comparable
to the effect of the LMC on the Galaxy and so is quite weak while if
$q_{tidal}\approx 8$, we might expect significant tidal torques
on the Galaxy that may cause interesting evolution in the form of
spirals, bars and warps.

\begin{figure}
\plotone{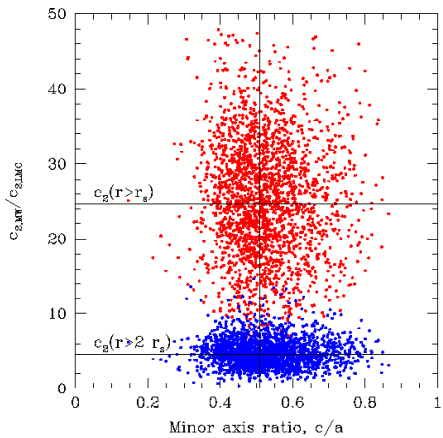}
\caption{{Relative values of the external quadrupolar tidal field of
triaxial dark halos measured by the parameter, $c_2$, for matter beyond
radius $r_s$ (red) and 2$r_s$ (blue). To give the value of $c_2$ a
physical meaning in the context of the Milky Way, we calculate $c_2$
for a satellite similar to the Large Magellanic Cloud with a mass,
$M=10^{10}$ M$_\odot$ at a distance $r=50$ kpc. We then calculate
$c_2$ for each dark halo and rescale the model to a Milky Way galaxy
model with mass $M_{200}=2\times 10^{12}$ M$_\odot$ and $r_s=26$ kpc.
The ratio $c_{2,MW}/c_{2,LMC}$ is therefore a measure of the relative
strength of the tidal field of the external quadrupole to a LMC
satellite. It is known that the strength of the LMC tidal field is
weak and has minimal dynamical effects but a field that is 5 to 10
times stronger can have a significant effect. The strength of the
external quadrupole field is 5 to 25 times the strength of the LMC for
matter beyond 1 to 2 scale radii suggesting interesting dynamical
phenomenology of the disk as the result of triaxial halos.}
\label{fig:c2}
}
\end{figure}

Figure~\ref{fig:c2} shows the distribution of values of the parameter
$q_{tidal}$ measured for the sample of halos using two different outer
radii, $r_0=r_s$ and $r_0=2 r_s$, plotted against the axis ratio
$c/a$. The halo axis ratios are determined using the normalized
inertia tensor of halo particles within $r < r_s$ using an iterative
method that gives a measure of the best fit ellipsoid of the
distribution \citep{dubinski_91}.  There appears to be no strong
correlation between the value of $c_2$ and halo shape since the
scatter is quite large.  The mean value of $q_{tidal}$ is about 5 for
$r_0=2 r_s$ compared to about 25 for a cutoff at $r_0=r_s$.  The
implication then is that a Milky Way disk that is misaligned with the
outer halo will experience a tidal field that is perhaps 5 to 25 times
the strength of the tidal field of an LMC-type satellite, depending on
the radial location where the misalignment begins. For halos with a
larger concentration $c$, both $r_s$ and $M_{200}$ are smaller for the
MW model but the values of $c_2$ vary within a factor of a few.  This
result is somewhat surprising for it implies that the external tidal
field acting on a disk galaxy from a triaxial halo is comparable in
strength to the tidal field of a large satellite having a close
encounter with a galactic disk.

\begin{figure}
\plotone{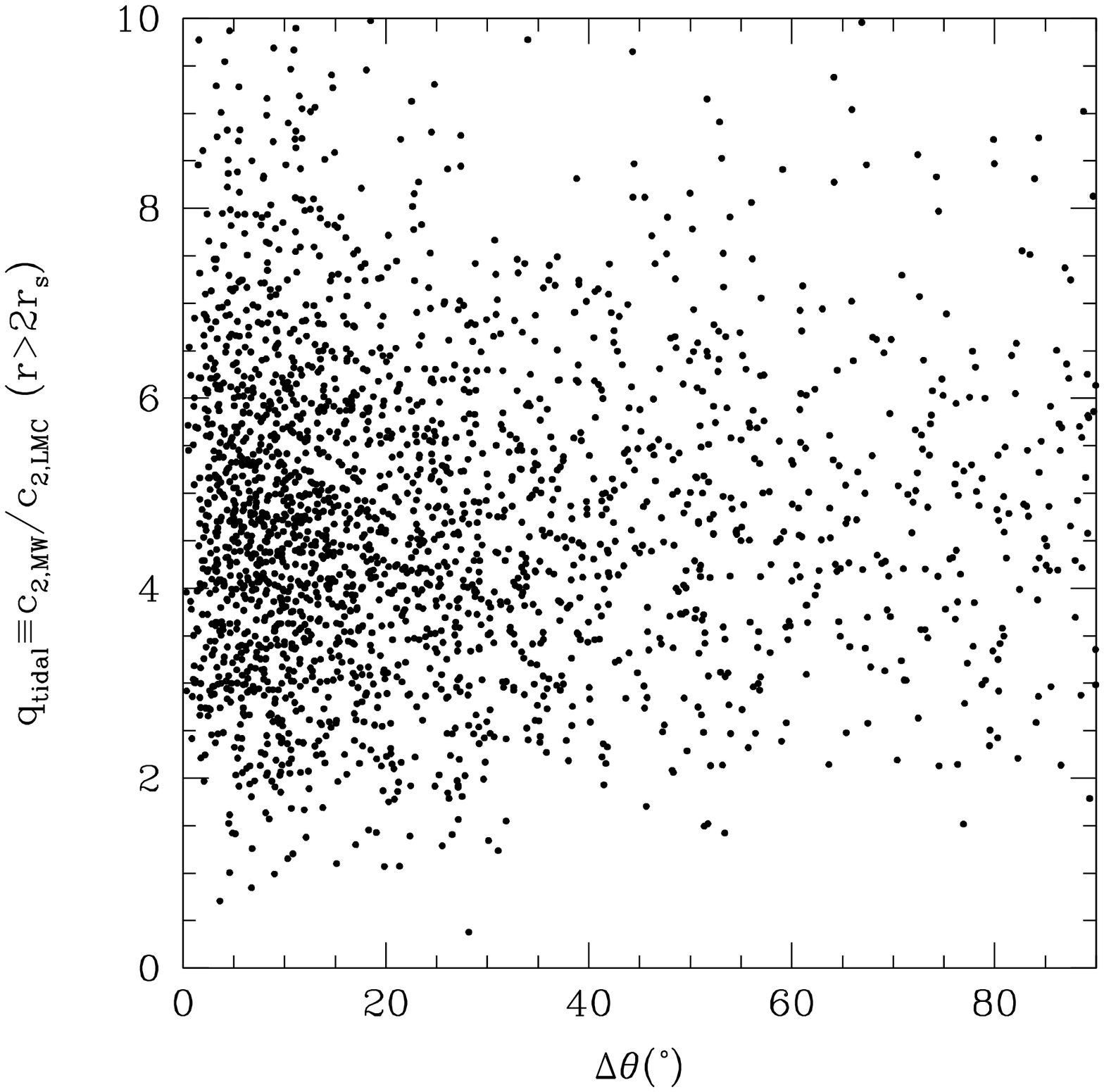}
\caption{
A scatter plot of the normalized strength of the external tidal field $q_{tidal}$
and the misalignment angle between the inner and outer halo $\delta \theta$ for 2200
dark halos.  
There is no correlation, suggesting that a significant fraction of galactic disks 
may experience relatively strong tidal torques from halo misalignment during their
evolution.
\label{fig-qtidal-align}
}
\end{figure}

Finally, the disk will only be torqued if there is a significant misalignment
with the halo independent of the strength of $q_{tidal}$. 
One might imagine that more flattened halos with the largest $q_{tidal}$ are
more closely aligned and thus have a weaker potential for torquing.   
We therefore looked for a correlation between $q_{tidal}$ and
the relative alignment of the inner and outer halo $\delta \theta$.
We see no such effect
(Fig.~\ref{fig-qtidal-align}) and conclude that significant
perturbations can arise from external torquing due to this misalignment.

We now go on to demonstrate some of these effects in controlled N-body simulations.

\section{Simulations}
\label{sec:simulations}
\noindent

We perform a series of $N$-body simulations using a modified version
of a parallelized treecode \citep{dub96}. The code has been changed to
include an external quadrupolar potential field that is tumbling at a fixed
pattern speed $\Omega_p$. Forces on particles are the sum of the $N$-body
force and the external potential field. If particles stray beyond a
radius $r_0$ then the field is shut off. For the simulations described
here, the radius is $r_0=60$ kpc. A model of a galactic disk plus
surrounding dark halo is simulated within this tidal field.  We also make
sure the model remains centered on the quadrupole field.  This prevents the
outer dark matter halo in the N-body model from being significantly
distorted.  

The model galaxies in question represent the Milky Way and Andromeda
(M31), as given by \citet{widrow_05} (WD herein).  These models both
have exponential disks with Hernquist-type bulges embedded within a
cuspy dark halo similar to an NFW profile.  We use the most stable
versions of these models called MWb and M31a in WD but we refer to
them simply as MW and M31 in this paper.  The M31 model has been
tested at a variety of resolutions and is known to be stable against
bar formation after run times of 10 Gyr \citep{gauthier_06}.  The MW
model on the other hand is mildly unstable and develops a bar later on
after $t=6$ Gyr as we shall see below.  The particles are distributed
with the following numbers in both the M 31 and MW runs:
$N_{disk}$=1,500,000, $N_{bulge}$=500,000 and $N_{halo}$=2,000,000.
The model parameters are provided in Table~1.

\begin{table*}
\caption{Parameters for the Milky Way (MW) and M31 models used in our
simulations (taken from Widrow \& Dubinski, 2005)
\label{tab-simulations}
}
\centering
\begin{tabular}{lcccccccccc}
\hline\hline
Model & $M_d$ & $R_d$ & $z_d$ & $r_s$ & $v_s$ & $v_b$ & $r_b$ & $R_0$ & $\delta{R}_0$ & $Q$\\
MW  & 3.3 & 2.81 & 0.44 & 8.8 & 345 & 435 & 0.88 & 30 & 1 & 1.3 at 2.5$R_d$\\
M31 & 7.7 & 5.58 & 0.60 & 12.9 & 337 & 461 & 1.83 & 30 & 1 & 1.25 at 2$R_d$\\
\hline
\end{tabular}
\begin{list}{}{}
\item Col. 1: Model for galaxy, Col. 2: Disk mass (10$^{10}$ M$_\odot$), Col. 3: Disk scale length (kpc), Col. 4: Disk scale height (kpc), Col. 5: NFW-halo scale radius (kpc), Col. 6: NFW-halo scale velocity (km s$^{-1}$), defined in Section~2, Col. 7: characteristic bulge velocity (km s$^{-1}$), Col. 8: bulge scale length, Col. 9: disk truncation radius (kpc), Col. 10: truncation width (kpc), Col. 11: Toomre $Q$-parameter. 
\end{list}
\end{table*}
We also populate the disks with particles out to large radii to better
investigate edge effects.  The MW model extends to 10 radial scale
lengths while the M31 model goes out to 7 scale lengths.  We note that
the scale radii of the MW and M31 halos are $r_s \approx 10$ kpc and
somewhat smaller than our putative MW halo in the discussion above.
The parameters are fitted values for observed rotation curves and
inferred brightness profiles of the two galaxies.  The {\em real}
inner halos of M31 and MW are likely contracted compared to the pure
dark matter models. The mass model fits to the real data have scale
radii that are smaller by a factor of two in accordance with
expectations of the contraction of halos by dissipative processes
during galaxy formation.

We have found it convenient to select a set of simulation
units that set the scale lengths and rotation velocities to values near
unity. We scale the simulation units to the physical units in the following 
way: 
\begin{itemize}
\item Gravitational constant $G$=1,
\item 1 simulation length unit = 4 kpc
\item 1 simulation velocity unit = 220 km s$^{-1}$,
\item 1 simulation mass unit = 4.5$\times10^{10}$ M$_\odot$,
\item 1 simulation time unit = 17.7 Myr
\end{itemize}

We express the general external tidal field as:
\begin{equation}
\Phi_{ext}(r,\theta,\phi) = \sum_l \sum_{m=0}^{m=l} P_{lm}(\cos\theta)
(a_{lm} \cos m\phi + b_{lm}\sin m\phi) r^l
\label{eqn:c2}
\end{equation}
The coefficients $a_{lm}$ and $b_{lm}$ describe the strength of the
field and are derived from Equation~\ref{eqn:coeffs}. If we assume that
the quadrupole is aligned with the principle axes and is due to a
plane-symmetric distribution then the $b_{lm}$ terms are zero.  For
these simulations we are only interested in the $l=2$ quadrupole
terms.
The pure external quadrupole field then becomes:
\begin{equation}
\Phi_{ext}(r,\theta,\phi) = 
\sum_{m=0}^{m=2} a_{2m} P_{2m}(\cos\theta) \cos(m\phi) r^2.
\label{eq-phi-ext}
\end{equation}

We estimate a suitable tidal field using the following procedure.
We first take an NFW profile that has been flattened into a perfect
ellipsoidal shape with axis ratios $q_1=b/a$ and $q_2=c/a$.
We then determine the spherically averaged density profile of this
distribution and fit a spherical NFW profile to determine the scale
radius, $r_s$ and mass $M_s$ as is done in cosmological simulations.
We are interested in the tidal field due to material beyond a radius,
$r_0= 2r_s$, so we compute the coefficients of the quadrupolar tidal field
for the material beyond this radius.  As an example, we find that for a
flattened NFW halo with $q_1=0.6$ and $q_2=0.5$ with a scale length
$r_s=26$ kpc and $M_{200}=2\times 10^{12}$~$\msol$ that the quadrupolar
tidal field coefficients in simulation units are:
\begin{itemize}
\item $a_{20}$ = 0.00050,
\item $a_{21}$ = 0.0,
\item $a_{22}$ = -0.00015.
\item The quadrupole amplitude is $c_2=0.00051$.
\end{itemize}

(If we used a more concentrated halo with $c=15$ then $r_s=16$~kpc and
$M_{200}=1.5\times 10^{12}~\msol$ leading to a field strength that
would be about 3 times larger for $r_0=2 r_s$.)  The N-body galaxy
models have a nearly spherical halo at large radii so we model the
effect of the triaxial halo by this quadrupolar field.  We assume that
the inner halo has aligned with the disk and is essentially
axisymmetric and remains aligned for the course of the simulation.  As
a comparison, the quadrupole coefficients for a LMC-like satellite of
$M=10^{10}$ M$_\odot$, at a distance of 50 kpc placed on the $x$-axis,
is $c_2=8.12\times 10^{-5}$ in simulation units so that $q_{tidal}=6.3$
close to expectation value for the cosmological distribution of halos
for material with $r>2r_s$..

To introduce a misalignment, the disk is tilted by 30$^\circ$ with
respect to the field and the field is made to rotate about the
$z$-axis with different pattern speeds. This choice of the tilt angle
is arbitrary but is approximately the expected angle for at least 30\%
of dark halos, according to the distribution shown in
Figure~\ref{fig-halo-align}.  The 4 pattern speeds that we choose are
$\Omega_p=0.11, 0.22, 0.44, $ and 0.88 $\kmskpc$ corresponding to
tumbling periods of $T=56,28,14,$ and 7 Gyr respectively.  We note
that these are very slow compared to the pattern speeds of barred
galaxies ($\Omega_p \approx 20~\kmskpc$) and so the corotation radii
are mainly larger than $r_{200}$ (The circular frequency at $r_{200}$
in the NFW model is $\Omega_{200} = v_{200}/r_{200} = H_0/100 \approx
0.7~\kmskpc$.)  These periods are approximately 4.0, 2.0, 1.0 and 0.5
times the age of the universe. We recall that \citet{bai04} find a
log-normal distribution peaking at $\Omega_p = 0.15 \kmskpc$,
corresponding to a tumbling period of 44 Gyr.  Thus, our simulations
probe the fast tumbling tail of this distribution -- the half of the
distribution that is most likely to yield interesting dynamical effects.  The
quadrupole coefficients and pattern speeds for the MW and M31 runs are
summarized in Table~2.
\begin{table*}
\caption{Simulation parameters \label{tab-quad}}
\centering
\begin{tabular}{rrrrr}
\hline \hline
Model & $a_{20} (10^{-4})$ & $a_{22} (10^{-4}$ & $\Omega_p$ & $T$ \\
MW-con & 0.0 & 0.0 & 0.0 & $\infty$ \\
MW-0 & 5.0 & -1.5 & 0.11 & 56 \\
MW-1 & 5.0 & -1.5 & 0.22 & 28 \\
MW-2 & 5.0 & -1.5 & 0.44 & 14 \\
MW-3 & 5.0 & -1.5 & 0.88 & 7 \\
MW-2a & 2.5& -0.75 & 0.44 & 28 \\
MW-2b & 1.3 &-0.38 & 0.44 & 28 \\
MW-2c & 0.63 & -0.19 & 0.44 & 28 \\
M31-con & 0.0 & 0.0 & 0.0 & $\infty$ \\
M31-0 & 5.0 & -1.5 & 0.11 & 56 \\
M31-1 & 5.0 & -1.5 & 0.22 & 28 \\
M31-2 & 5.0 & -1.5 & 0.44 & 14 \\
M31-3 & 5.0 & -1.5 & 0.88 & 7  \\
\hline
\end{tabular}
\begin{list}{}{}
\item Col~1: model name, Col~2: component $a_{20}$ of the external quadrupolar field (10$^{-4}$ M$_\odot$s$^{-2}$), Col~3: component $a_{22}$ of the external quadrupolar field (10$^{-4}$ M$_\odot$s$^{-2}$), Col~4: halo pattern speed $\Omega_p$ (km s$^{-1}$ kpc$^{-1}$), Col~5: tumbling period (Gyrs)
\end{list}
\end{table*}

We simulate each run for 8000 equal timesteps with $\delta t=0.05$ units, 
corresponding to about 7.1 Gyr in physical units. We soften
gravity with a Plummer softening kernel with radius of 40 pc for
the stars and 100 pc for the dark matter. Energy is
conserved typically to within 0.5\% and total angular momentum to within
1\%.

\section{Results}

\begin{figure*}
\plotone{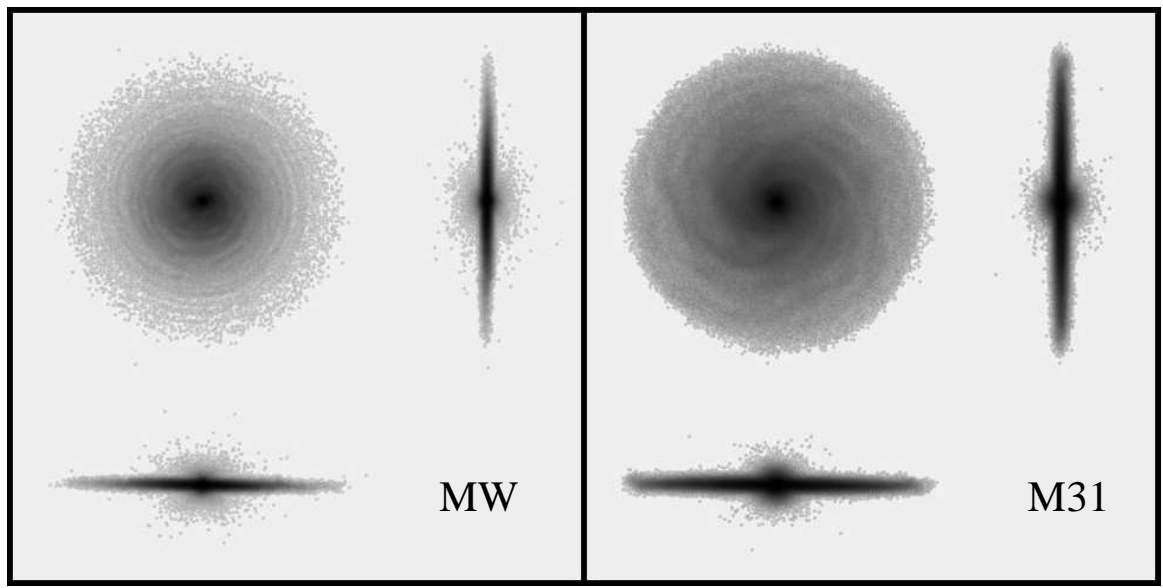}
\caption{{The state of the Milky Way and M31
disks (left and right respectively) at the end of control 
simulations with no external tidal potential.
While a very weak oval
distortion can be noticed in the Milky Way disk at this time
($\approx$ 6.4 Gyrs), the M~31 disk does not show any signature of a
bar. The warping of the disks is also absent and the disks remain in their initial plane.}}
\label{fig:control}
\end{figure*}

Before exploring the results for externally torqued disks, we first 
examine the evolution of control models.  Figure \ref{fig:control} shows 
the final state for the MW and M~31 models in the absence of any external 
torquing potential.  The disks remain co-planar and there are minimal 
signs of disk thickening and warping resulting from the amplification of 
the Poisson noise in the N-body disk.  
There are some remnants of spiral structure in the disks which arise from swing
amplification of the noise in the disk.  This transient spiral structure 
heats the disk azimuthally and slowly fades away as the simulations progress.

\begin{figure*}
\plotone{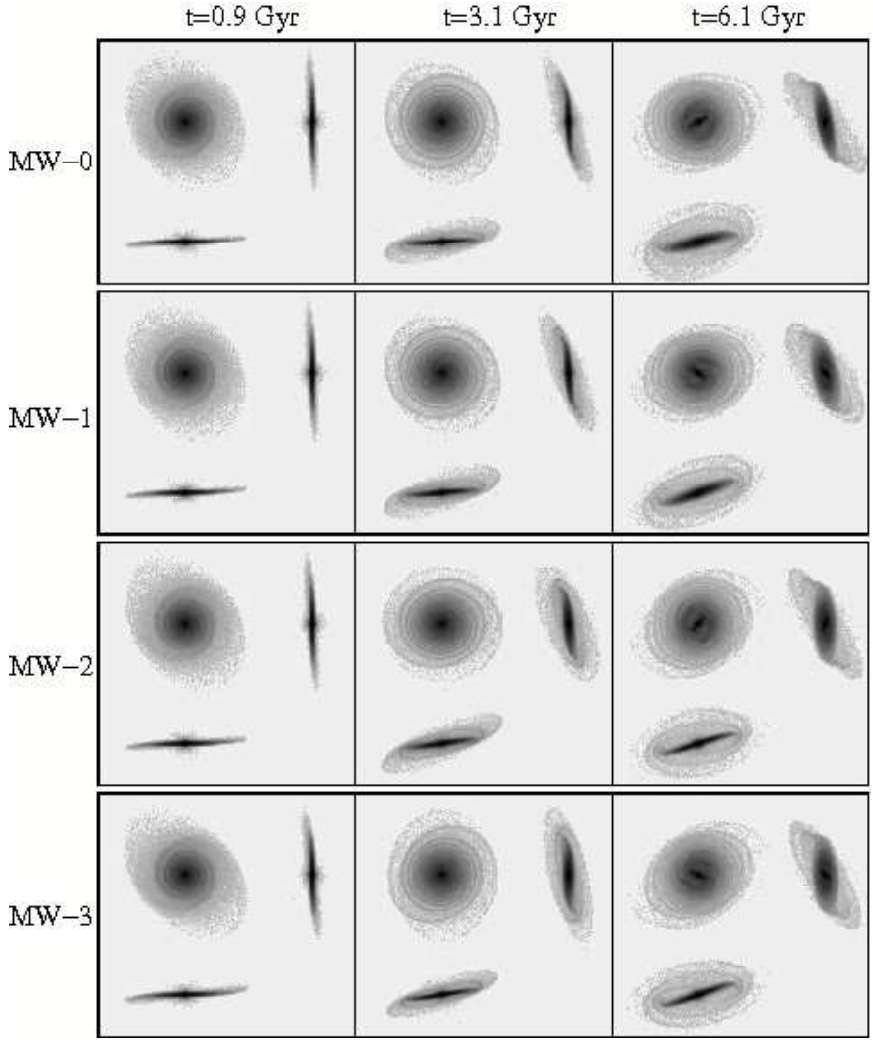}
\caption{{Snapshots of the Milky Way disk, at three different
times from the four different runs near the beginning, middle and end of the
simulation. The three views of the disk are
presented from the point of views that are initially face-on ($x-y$ top-left) and the
two perpendicular edge-on views ($x-z$ bottom-left; $y-z$ top-right).  
The disk precesses in all cases and the tilting induces a transient warp in 
the outer regions of the disk that persists for about 7 Gyr. The tidal forcing also
excites spiral structure in the outer disk.
\label{fig:mw_snap}
}
}
\end{figure*}

\begin{figure*}
\plotone{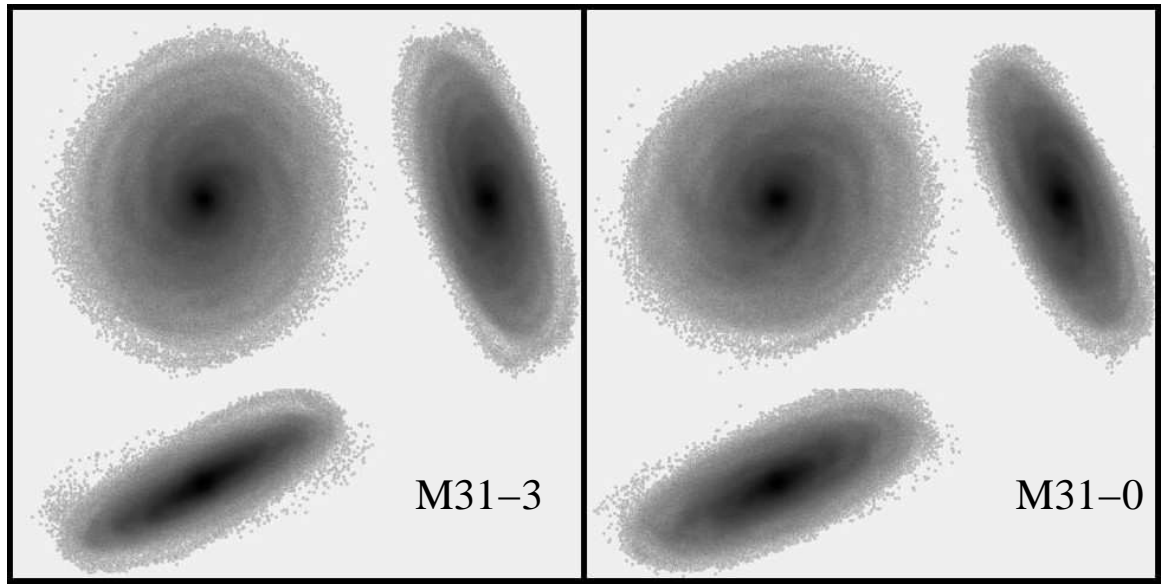}
\caption{Final snapshots of two representative M~31 runs shown at $t=7.1$ Gyr.
The left panel corresponds to the three views of the M~31 disk (with the same layout
as Fig.~\ref{fig:mw_snap}) at the end of the run
M31-3 with the fastest forcing pattern speed. The panel on the right shows the end of run
M31-0 with the slowest pattern speed.  The warping of the disk is more subtle in this
case but is still apparent.  
\label{fig:m31}}
\end{figure*}

\begin{figure*}
\plottwo{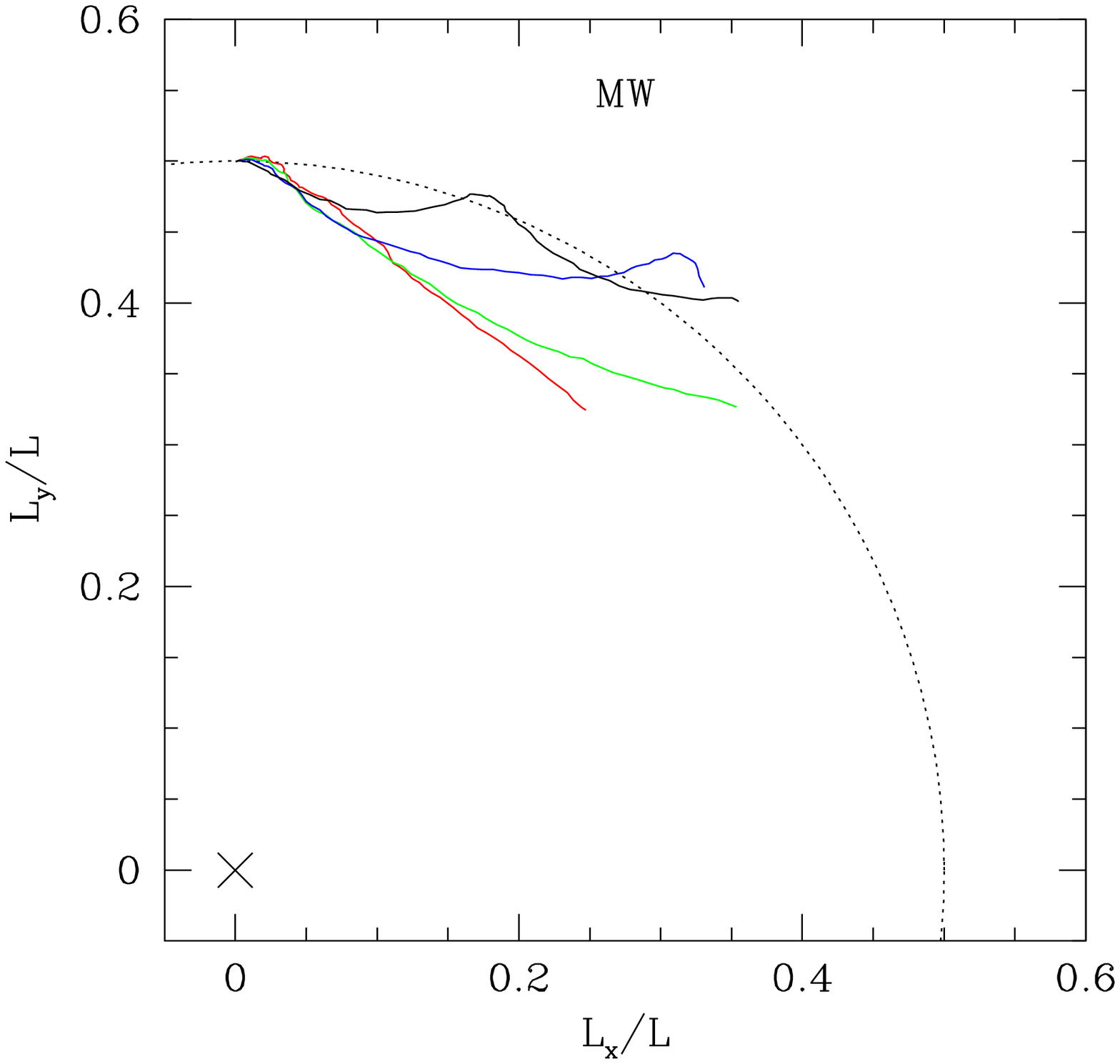}{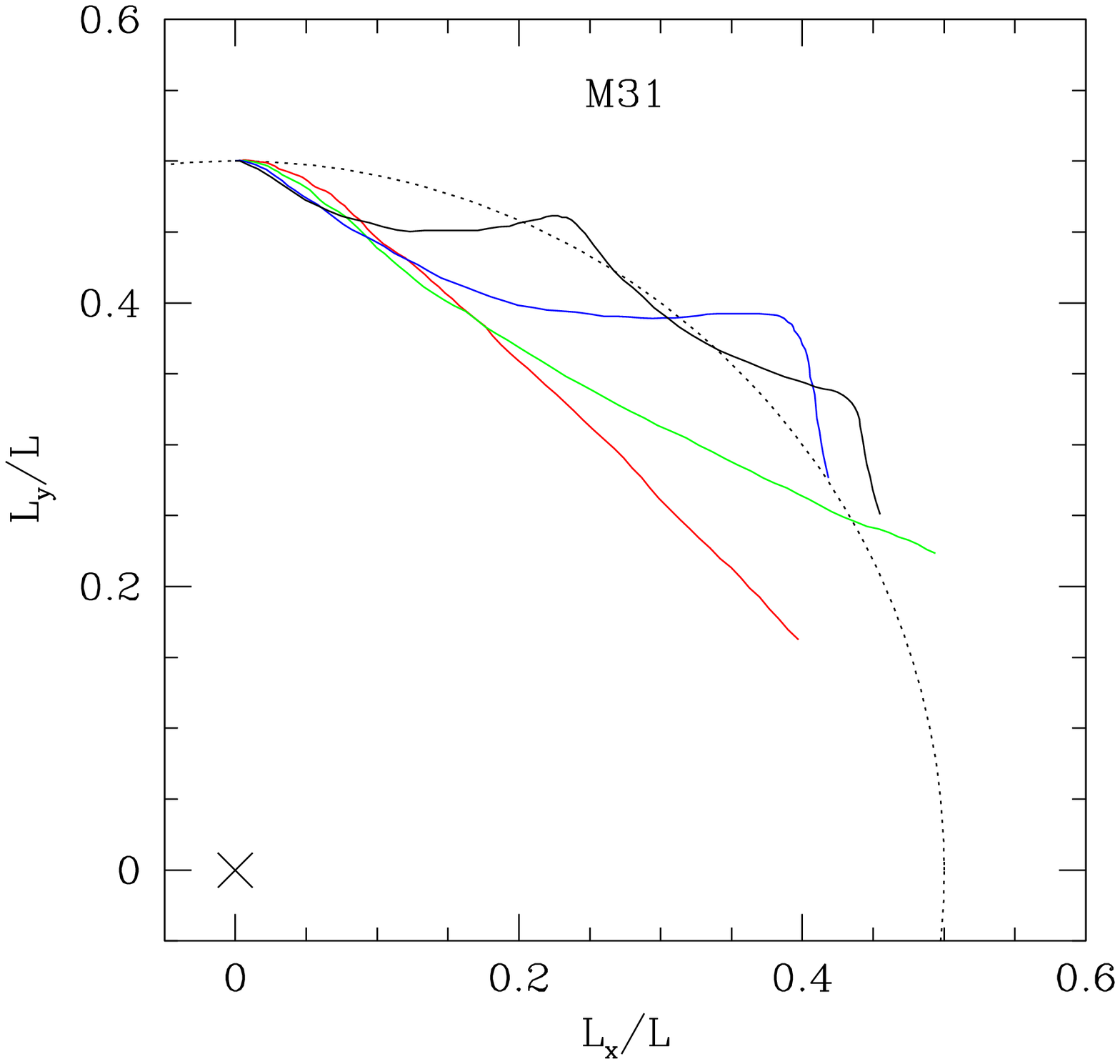}
\caption{The evolution of spin axis of the MW and M31 models - MW-0, M31-0
(red); MW-1, M31-1 (green); MW-2, MW-3 (blue); and MW-3, M31-3 (black). 
This time evolution is represented here by the normalized components of
the disk angular momentum vector $L_x/L$ and $L_y/L$.
The circle corresponds to a fixed inclination of $\theta=30^\circ$.  
The tumbling axis of the quadrupole potential
is perpendicular to the plane of this plot.  The disk responds by precessing in the
counter-clockwise direction as expected for this quadrupole potential.  
Nutations are clearly seen as the disk responds
to the triaxial halo.  The nutational period is generally smaller than the
precession frequency.  Even after the 7 Gyr time of the simulation the disk
only precesses through about 60 degrees.
\label{fig-inc}
}
\end{figure*}

When we turn on the external quadrupole potential with the expected 
amplitudes from cosmological dark halos,  we clearly see the various 
effects predicted for disk torquing.
Figures \ref{fig:mw_snap} and \ref{fig:m31} show the 
evolution of representative models of the MW and M 31 runs from 3
perpendicular views.
Associated videos\footnote{Animations are available
at the website http://www.cita.utoronto.ca/$\sim$dubinski/warpmovies}
also show how the external torque causes the
disk to precess and nutate like a tilted gyroscope on a table top.  
The evolution of the direction of the spin vector of the disks described by the
normalized $x$ and $y$ components of the angular momentum quantifies
this behavior (Fig.~\ref{fig-inc}) and the nutational frequency depends 
on the pattern speed of the forcing quadrupole potential. 
All runs indicate that the model disk precesses through an angle of
30-60$^{\circ}$ over 7 Gyr as it responds
to the external quadrupolar field implying precession periods of more
than $40$ Gyr or precession rates less than 0.15 rad Gyr$^{-1}$.  
We see below that these precession rates are in accord 
with analytical estimates from a rigid disk model forced by an external
potential.  These precession rates are about an order of magnitude smaller
than those found in models with a disk misaligned within a
triaxial potential \citep[e.g.,][]{kui91,jeo09} used in 
models of warping behavior that postulate that the main source of torque on
a disk is due to he potential of the {\em inner} part of the triaxial halo.
The phenomenology that we are exploring is therefore different
and the resulting effects are slower and more subtle. 
So while the precession
periods are long compared to the Hubble time, a disk can slew through
and a significant angle over its lifetime and this dynamical evolution
leads to a more gradual transient warping at the disk edge in a mechanism
similar to those invoking the accretion of angular momentum
\citep{ostriker89}.
We conclude that this must be an
important process in disk dynamical evolution over the age of the
universe.  For real galaxies, we expect a range of halo tumbling pattern
speeds, tidal field strengths and initial tilt angles so there should be
also be a range of warping behavior in the population at large.

\subsection{Warps}
\begin{figure*}
\plotone{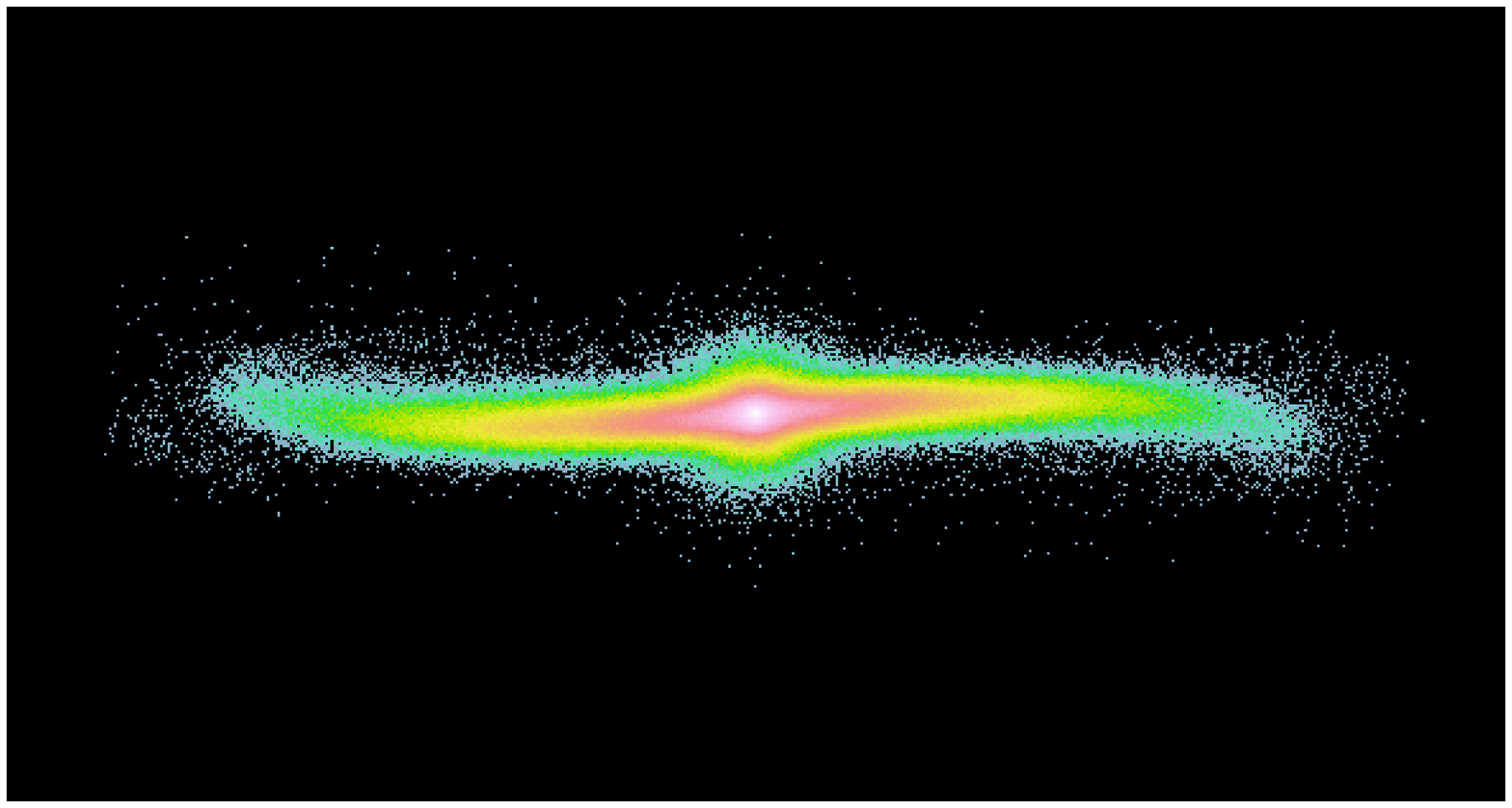}
\caption{
A prominent warp develops by the end of the simulation in model M31-3 with
a quadrupole field pattern speed of $\Omega_p=0.88 \kmskpc$.
\label{fig-warp}
}
\end{figure*}

\noindent
While the the disk stars within approximately 5 radial scale-lengths
stay within the plane and act like a rigid body because of their
self-gravity, stars near the edge of the disk begin to precess at different
frequencies and begin to warp away from the disk.  
Our simulations indicate that the model disks develop strong warping
early on in the runs and that such warping is sustained through the
length of the run (about 7 Gyrs) (Fig.~\ref{fig-warp}). 

The disk demonstrates strong flocculent spiral pattern, to even the
disk-edge where self-gravity is considered too weak to excite such
structure.  This is advanced as the handiwork of the external
quadrupole since we do not notice prominent outer spiral patterns in
the control runs. This is particularly noticeable in the
MW runs that have a more extended disk.

Following \citet{lev06}, we quantify the disk warps that develop in the
simulations by analyzing 
the vertical deviation of the disk from a plane at different radii
through the function,
\begin{equation}
W(\phi) = \sum_m W_m e^{im\phi}
\end{equation}
where $\phi$ is the particle azimuth. 

The inner, nearly co-planar disk is considered to be in 
the $x-y$ plane. The particles
inside the $j^{th}$ ring - of radius $R_j$ and width $\delta R$ - are
assigned the height $z_j$ from the midplane with $\phi_j$ defined as the
corresponding particle azimuth. We first rotate into the frame of the
inner disk before the analysis. We determine the midplane
of the disk by computing the moment of inertia of the disk particles
within 3 scale lengths where the disk is nearly planar and then
diagonalizing the inertia tensor to give the orientation of the disk.
We determine the coefficients $W_{m}$ using a Fourier analysis of the particle
distribution through,
\begin{equation}
W_{m} = \frac{1}{N_k} \sum_j z_j e^{-im\phi_j}
\end{equation}
The amplitude of the deviation is given by $|W_m|$ and
the phase angle through $\phi=\tan^{-1}[Im(W_m)/Re(W_m)]/m$.
We can then construct functions of $W_m$ versus $R$ to examine disk warping.
The function $W_0(R)$ corresponds to $m=0$ bowl-shaped deviations, $W_1(R)$
corresponds to the usual $m=1$ integral-sign shaped warps while $W_2(R)$ are
second order ``scalloped" warps.

\begin{figure*}
\plotone{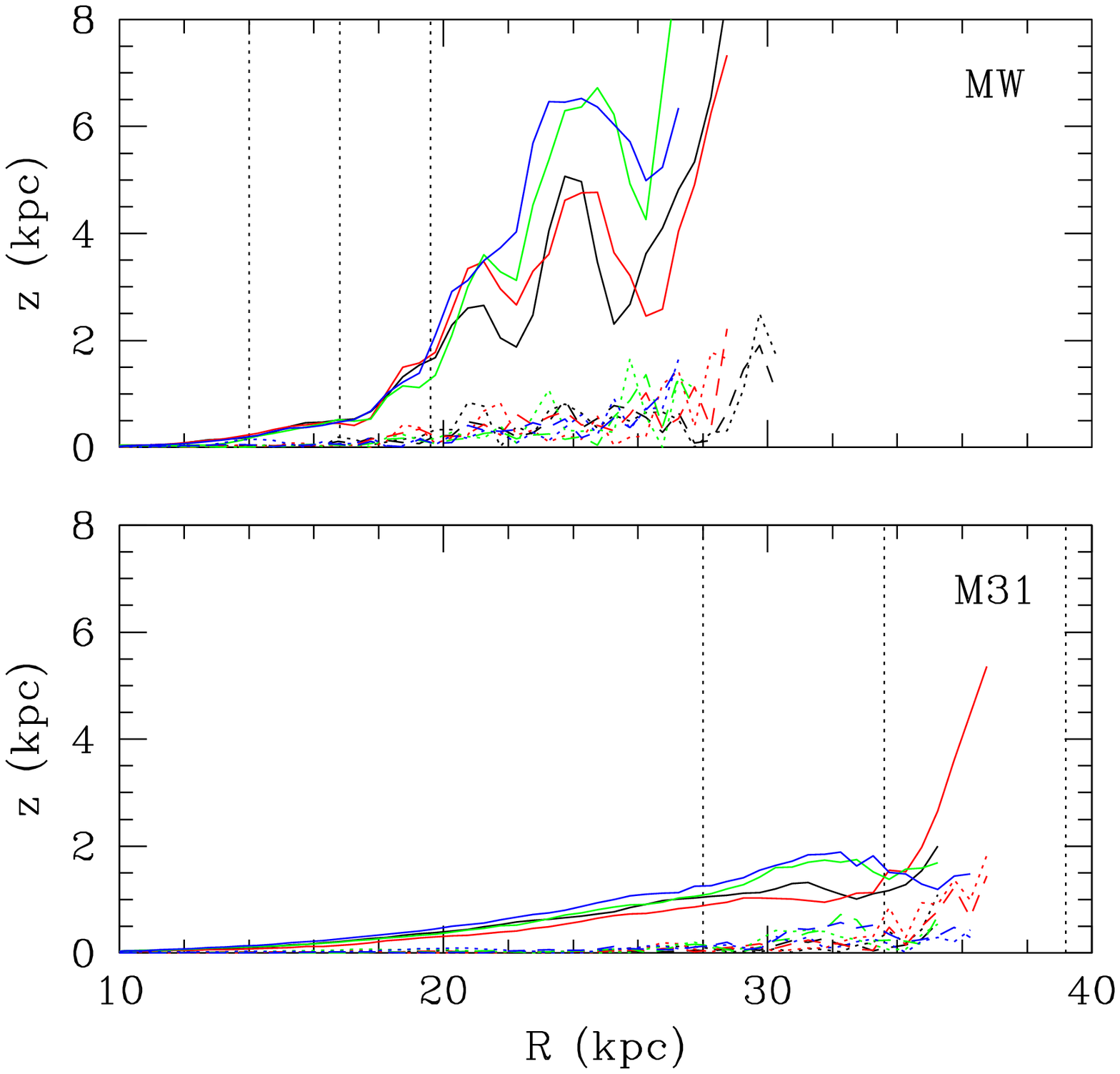}
\caption{
Disk warping behavior quantified by the functions $W_0(r)$ (dotted lines),
$W_1(r)$ (solid lines), and $W_2(r)$ (dashed lines) for the four MW and M31
runs with halo tumbling periods of $P=56, 28, 14,$ and 7 Gyr (black, red,
green and blue) respectively at the last snapshot at $t=7$ Gyr.  
Disks generically develop vertical
deviations over the radial range of 5, 6, and 7 $R_d$ 
(vertical dashed lines)  and
are dominated by the $m=1$ integral-signed warp with stars reaching 2-5 kpc
above the plane depending on the model.  The quantitative behavior of these
warps are similar to that seen in real galaxies including the Milky-Way and
M 31.
\label{fig-w1}
}
\end{figure*}

Figure \ref{fig-w1} shows the results of this analysis for the 4 models
of both the MW and M31 system.  The dominant vertical deviation 
is the $m=1$ integral-signed warp.  For the MW models, the warp turns upwards
at $R\approx 15$ kpc and continues to $R=30$ kpc reaching about 5
kpc above the plane.  The M 31 models shows a qualitatively similar
behavior though the warping begins at the larger radius of $R\approx 20$
kpc but only reaches about 3 kpc above the plane at a radius of $R=40$ kpc.
The M31 model is more extended and massive than the MW model while being
forced by the same external tidal field and so shows a smaller degree of
warping behavior.

\begin{figure*}
\plottwo{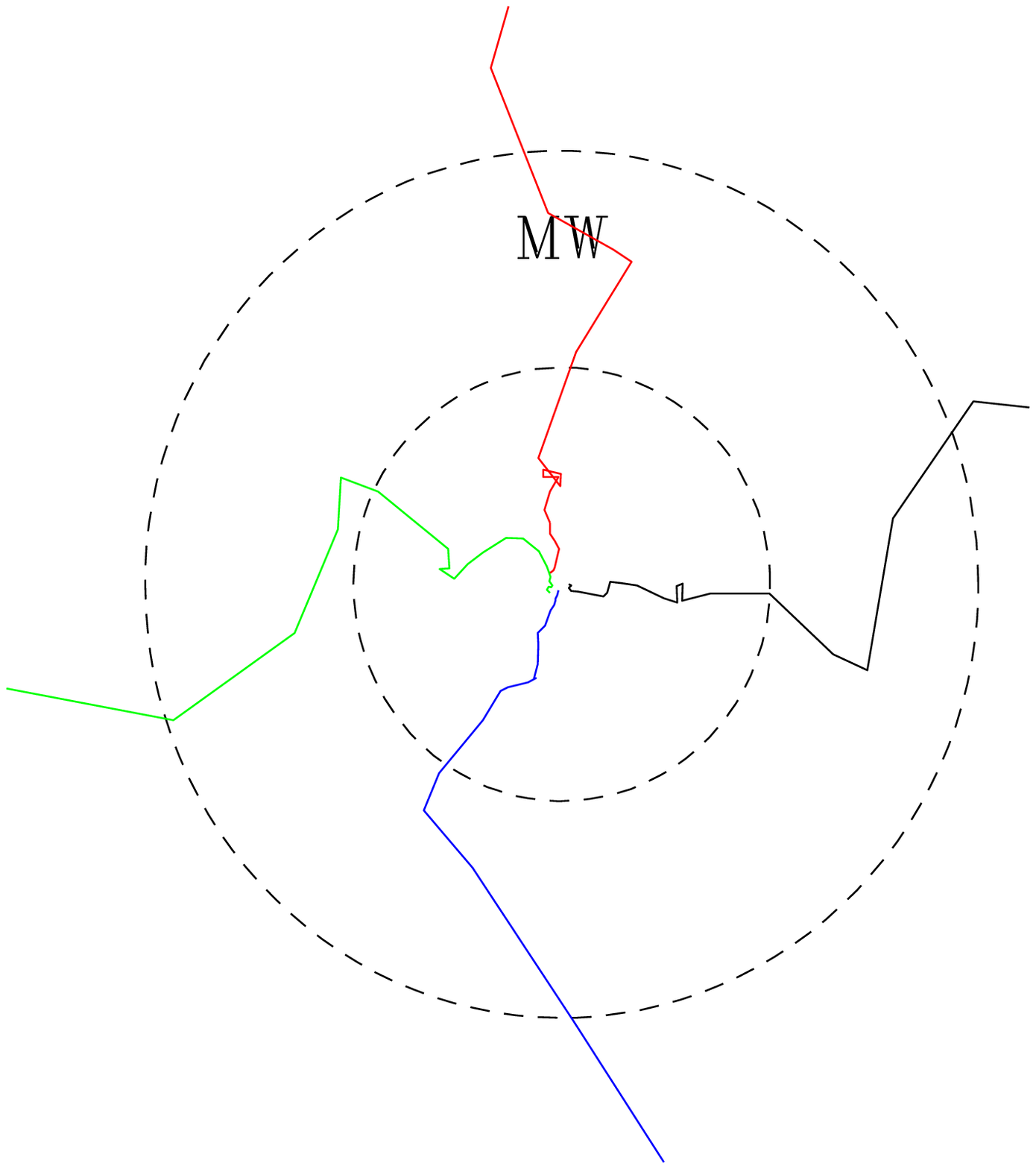}{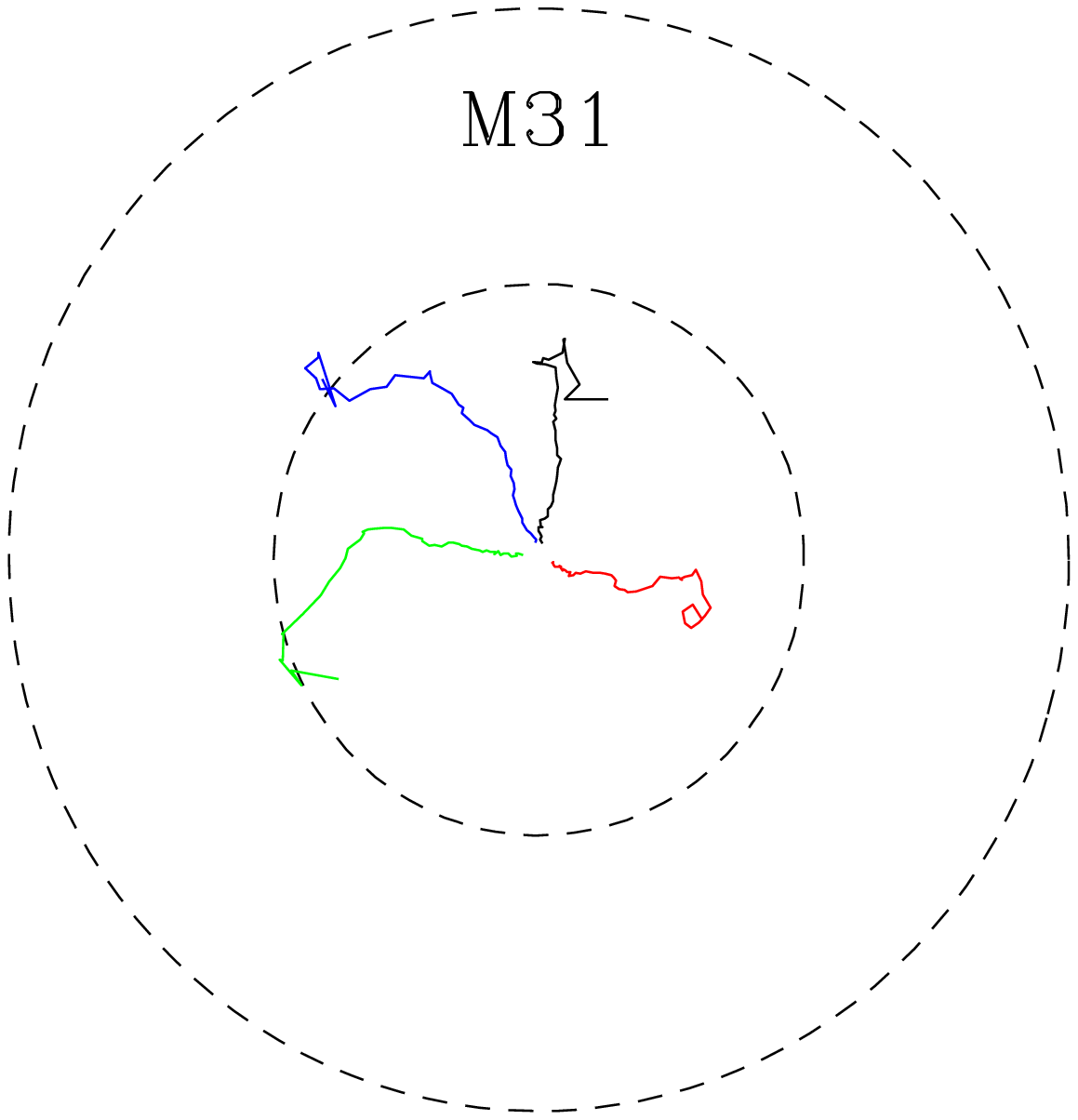}
\caption{The inclination (tip) versus longitude or tip-LON plots
for the warps in the MW disk models at the last snapshot of the simulations 
at $t=7$ Gyr.  
The four models with tumbling periods of $P=56, 28, 14,$ and
7 Gyr are give by the black, red, green, and blue lines respectively.
The dashed circles correspond to 3 and 6 degrees of tip
respectively while the longitude is measured around the circles.  The line
of nodes of the warp is approximately straight when it begins to move out
of the plane but becomes more erratic at higher inclinations.  The line of
nodes tends to curve in a leading spiral form similar to what is observed
in real warps.
The warps are less pronounced in the M31 models but still behave in a
similar manner.
\label{fig-tiplon}
}
\end{figure*}

We also examined the alignment of the warps by measuring the line of nodes
(LON) from the phase angle determined from the vertical Fourier analysis.  
We present the tip-LON plots for our models
following \citet{bri90} (Fig.~\ref{fig-tiplon}).  
The warps begin with a relatively
straight line of nodes when they begin to move out of the plane
at the disk edge but generally show a
leading spiral curving out to $\sim 6 R_d$ following the observations from 
Brigg's analysis.  At larger radii, the warping becomes less coherent
resulting mainly by the onset of differential precession of the stellar 
disk orbits and the modulated forcing of the tumbling external quadrupole
potential.  While we are only modeling a pure N-body system one
expects that the mechanism will generate coherent warping in the gas as
well at least out to 6 scale lengths.

\subsection{Bars}

\begin{figure}
\plotone{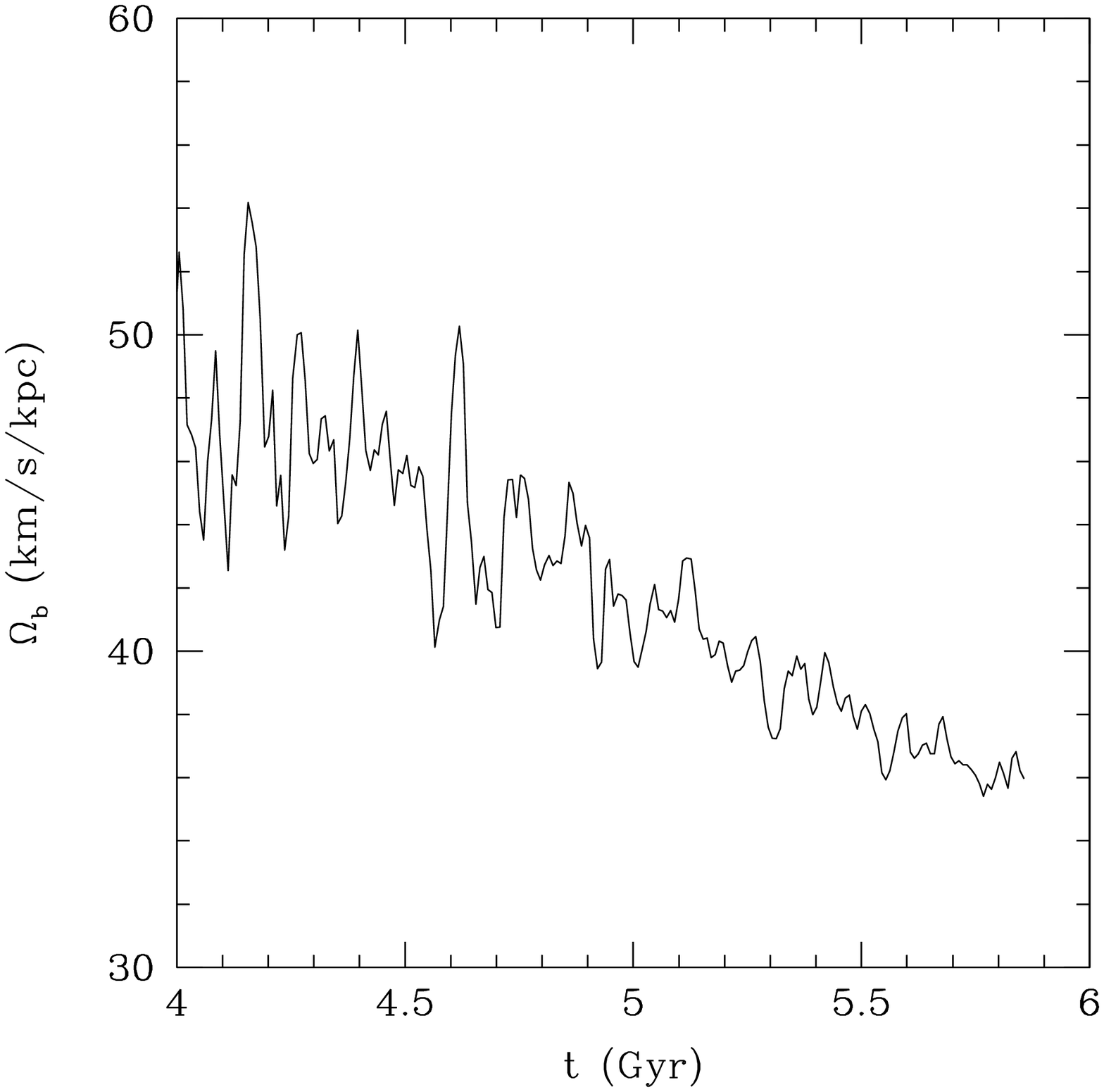}
\caption{
Pattern speed evolution for model MW-2 from $t=4$ Gyr.  The disk precesses
initially in response to the external torques from the quadrupolar field
and at late times develops a bar instability. The bar pattern speed decays
in the usual manner in response to resonant interactions with the dark
halo.
\label{fig-pattern}
}
\end{figure}

The tidal distortion due to a galaxy interaction in a fly-by 
can trigger the bar instability under some conditions
\citep[e.g.][]{nog87,ger90}.  The effect is due to a transient external
quadrupole that disturbs the stellar orbits in the center of the galaxy.  
Under our hypothesis of a tumbling misaligned halo, there is also
a changing external quadrupole potential that can perturb the disk and
potentially trigger the bar.  The physical situation is somewhat different
than a impulsive galactic fly-by however, in that rate of change of the
orientation is slow and the amplitude of the tidal potential is constant.
Nevertheless, the Milky Way disk is found to develop a fast bar 
at about 3 Gyrs.  Once formed, 
the bar is sustained and its pattern speed declines
from an initial value of 50 to 35 $\kmskpc$
by the end of the run in accord with typical expectations for models
like this (Fig~\ref{fig-pattern}) \citep[e.g.,][]{deb98,one03,dub09}.

\begin{figure}
\plotone{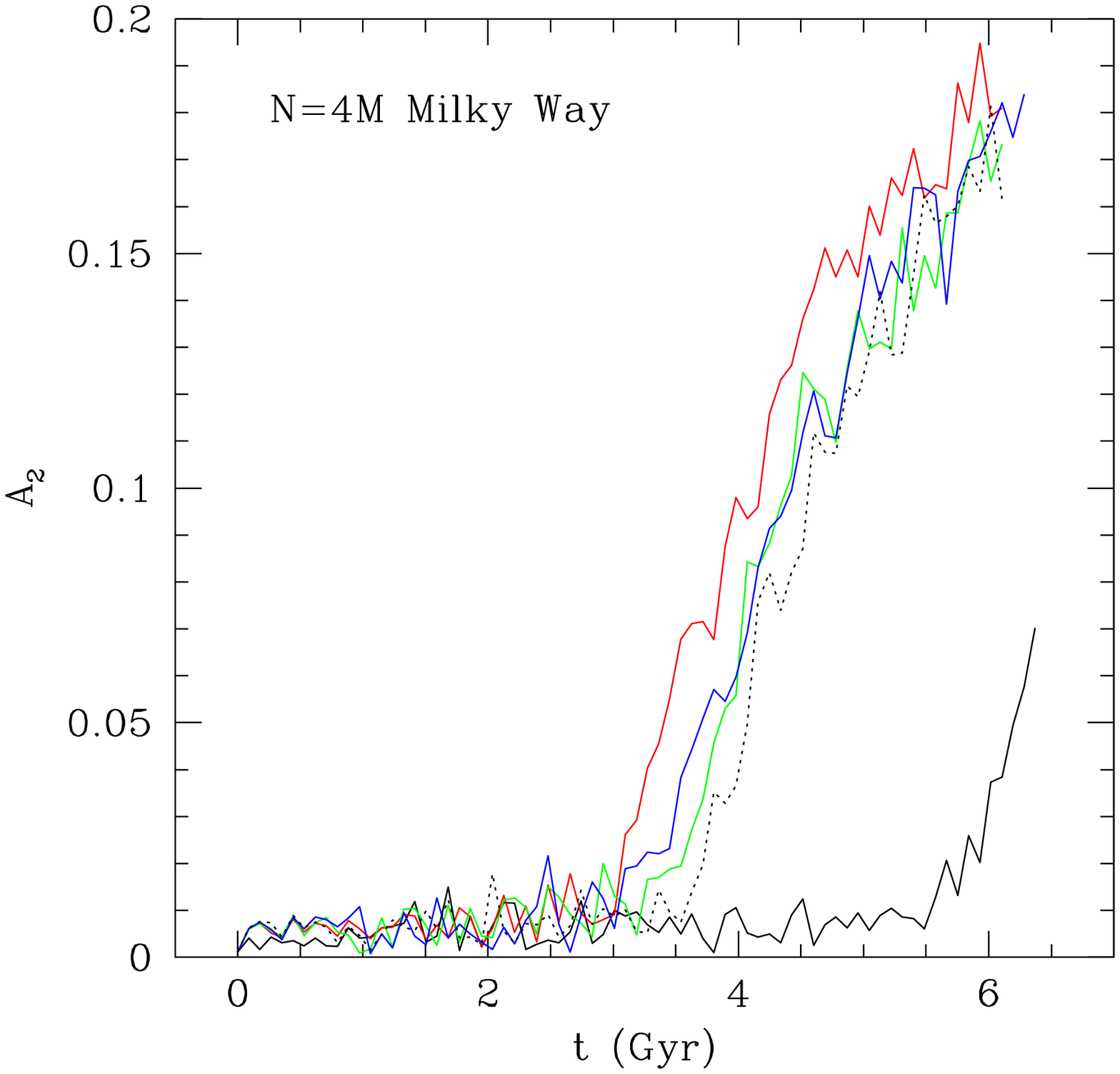}
\caption{{The evolution of the bar strength $A_2$ is shown here
for the 4 Milky Way runs as well as for the control run done with the
Milky way model. Here, the bar strength is measured in terms of the
quadrupolar strength parameter of the bar that forms in the disk. The
broken black line corresponds to the fastest halo ($MW-3$), while
the green curve refers to $MW-2$, followed by the blue and red
curves, which depict $A_2$ evolution for $MW-1$ and $MW-0$. The
control run is represented by the solid black line.}}
\label{fig:a2}
\end{figure}

Figure~\ref{fig:a2} shows the evolution of the strength of the bar
that is triggered in the Milky Way for different models
by the torquing provided by the
external halo. We find a well-developed bar, of nearly similar
strengths, forming by about 3 Gyrs in all the runs, indicating that
the rate of tumbling of the halo is immaterial to the bar
strength. However, $\Omega_p$ of the quadrupole  appears to have a weak 
effect on the time of onset of the bar instability - this is hastened with
increasing tumbling period $T$, i.e. falling pattern speed of the
external halo, though this is not marked; the fractional difference in
bar onset times between the runs done with the slowest and fastest
halo in the Milky Way model ($MW-0$ and $MW-3$, respectively) is
about 16$\%$. This small difference in the bar onset time is also
noticed in the snapshots from these two runs, shown in
Figure~\ref{fig:mw_snap}.

Instead, we find that it is the strength of the quadrupole of the
external halo that is the crucial factor in controlling the bar
instability. In fact, a control run performed with a null external
quadrupole leads to a much later ($\approx$ 5.8 Gyrs) onset of a very
weak bar in the Milky Way disk (maximum bar strength is less than 0.4
times the bar strength reached in the other runs); see
Figure~\ref{fig:control}. Thus, we assure ourselves that the inherent
bar unstable nature of the Milky Way model is very weak, albeit
non-zero. Additionally, the control runs for both models result in
significantly weaker spirality, especially at the edges of the disk and
no warp.  

The M~31 disk is found to be unaffected by the halo torquing, in
regard to bar formation. This is apparent from the pictures of the
M~31 disk at the end of the simulations (Figure~\ref{fig:m31}); the
result is found independent of the halo pattern speed used in the run.

\section{Analytic Rigid Disk Model}
\noindent
It is interesting to gauge the effects of the external
quadrupolar term in the potential of a tumbling triaxial dark halo in an
analytical model using
a rigid, thin exponential disk for comparison to the phenomenology in
the simulations.  We emphasize again that the inner disk and halo are
tightly coupled through dynamical friction and so remain aligned.  The
only possible source of torquing on the disk is a misaligned and
possibly tumbling outer halo.  The assumption of a rigid disk is a
reasonable assumption since the simulations show that the disk
maintains coherence because of self-gravity within $R\la 5 R_d$ even
while it tips in response to the external tidal torque.  We use the
classical Euler equations of motions for a rigid spinning disk.  The
dynamic variables are the standard Euler angle
($\theta,\:\phi,\:\psi$) where $\theta$ is the inclination of the
disk, $\phi$ is the longitude of ascending node and $\psi$ the angle
about the disk's symmetry axis.  In the presence of a time dependent
torque, the disk will precess and nutate and so $\theta$ and $\phi$
will evolve in time.  In these calculations, we set up the disk with
an initial inclination of $\theta=30^\circ$ and set the halo
quadrupole potential tumbling about the $z$-axis with $\theta=0$ for
direct comparison to the results of the simulations..

\subsection{Euler Equations of Motion}
\label{sec:eom}
\noindent
After \citet{goldstein} (see p.210  and equation 5.52),
the Lagrangian
of the disk is
\begin{equation}
{\cal L} = \displaystyle{
			\frac{I_1}{2} 
			(\dot{{\theta}^2} + \dot{{\phi^2}}\sin^2\theta) +
			\frac{I_3}{2}
			(\dot\phi\cos\theta + \dot\psi)^2 -
			V(\theta, \phi, t)
			},
\label{eqn:lagrangian}
\end{equation}
where $I_3$ is the moment of inertia about the symmetry axis
$I_1=I_2$ are the moments measured about an axis in the plane of the disk.
The potential energy $V$ is determined from the interaction between a rigid
thin exponential disk and a tumbling external quadrupole potential.
The equations of motion can then be written as:
\begin{eqnarray}
I_1\ddot{\theta} - I_1\dot{\phi}^2\sin{\theta}\cos{\theta} +
S\dot{\phi}\sin{\theta}  + \displaystyle{\frac{\partial{V}}{\partial\theta}} 
&= 0\\
I_1\ddot{\phi}\sin^2{\theta} + 
2I_1\dot{\phi}\dot{\theta}\sin{\theta}\cos{\theta} -
S\dot{\theta}\sin{\theta} + 
\displaystyle{\frac{\partial{V}}{\partial\phi}} &= 0,
\label{eqn:eoms}
\end{eqnarray}
where $S= I_3(\dot{\phi}\cos{\theta} +\dot\psi)$ (the disk angular
momentum) is a constant of motion and $V$ is the potential energy
between the disk and the tumbling halo. A tilted disk will naturally
precess.  For example, with an axisymmetric potential $V(\theta)$, the
expected precession frequency can be derived from the first Euler equation
by assuming that $\theta$ is constant.
For small $\dot{\phi}$ we expect
\begin{equation}
\dot{\phi}_{prec} \approx -\frac{1}{S \sin\theta} \frac{\partial V}{\partial\theta}
\label{eq-prec}
\end{equation}
\citep[cf.][]{goldstein}.
If the potential also depends on $\phi$ and is rotating about the $z$-axis, 
we can expect a more complex evolution combining both precession and nutation.

\subsubsection{Direct Solution of the Euler Equations for a Rigid Exponential Disk}

We first consider the interaction of a rigid exponential disk with
an external quadrupolar tidal field for comparison to the results 
of the N-body simulations.  We point out that this process is quite 
different from the disk precession examined in work that has treated 
a triaxial halo as a rigid background potential \citep{kui91,jeo09}.
In those treatments, the precession rate is an order of magnitude larger
due to the much larger torques felt by the inner halo.  However, the large
torque in these simulations is artificially high since we expect dynamical
friction to align the disk and inner halo within a few disk dynamical times
\citep{dubinski_kuijken}.
In this study,
we assume that only the outer halo
can exert any torque and subsequently the strength of the torque is 
considerably smaller, typically by an order of magnitude.

In Appendix A, we compute the time dependent interaction potential
between a rigid exponential disk and an external quadrupole potential
to plug into the Euler equations. The potential is:
\begin{eqnarray}
V(\theta,\phi; t) &=& 3 M_d R_d^2 \left \{ -\frac{a_{20}}{2}(3\cos^2\theta - 1) 
\right . \nonumber \\
&& \left . + 3a_{22}\cos[2(\Omega_p t - \phi)] \sin^2\theta \right \}.
\label{eq-pot}
\end{eqnarray}
where $M_d$ and $R_d$ are the mass and exponential scale radius of the disk,
$a_{20}$ and $a_{22}$ are the quadrupole terms for the external terms of a
triaxial potential in equation \ref{eq-phi-ext}, $\Omega_p$ is the 
tumbling frequency of the quadrupole potential assumed to be about 
the $z$-axis with $\theta=0$ and we assume that $\phi=0$ at $t=0$.
We can compute the moments of inertia $I_1$ and $I_3$ and the spin $S$ 
directly from the N-body model and so fully specify the Euler equations 
for our case.    

\begin{figure}
\plotone{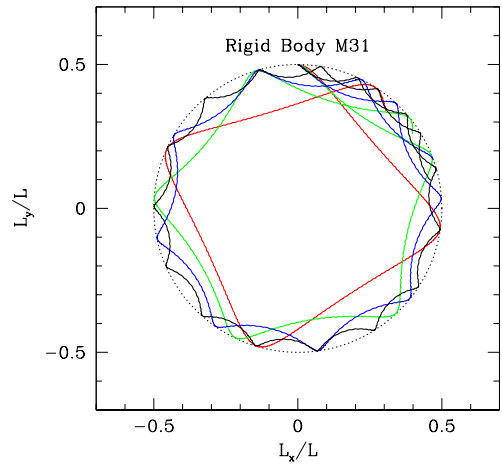}
\caption{{The trace of the spin axis through the normalized
components of the disk angular momentum vector $L_x/L$ and $L_y/L$
for the rigid body model
of M31 forced by an external quadrupole potential.  Colored lines correspond 
to the same pattern speeds applied to N-body models with M31-0 (red); 
M31-1 (green); M31-2 (blue); and M31-3 (black). The agreement with 
Fig.~\ref{fig-inc} is very good suggesting that self-gravitating disks 
behave like rigid bodies in this context.  However, different pattern speeds 
lead to different nutational frequencies and so the edges of disks may flop 
around at different rates.
}
}
\label{fig-xi-vs-eta}
\end{figure}

For the M31 model, we solve the Euler equations of motion numerically
using the same initial state as the N-body model along with four
different pattern speeds for the quadrupole potential.  We use the
values of $I_1=3 M_d R_d^2$ and $I_3=6 M_d R_d^2$ expected for an
exponential disk as well as the spin $S$ (disk angular momentum)
measured from the N-body models.  The specific parameters for the
M31 model in dimensionless units are $M=1.75$, $R_d=1.39$ so that
$I_1=10.17$, $I_3=20.34$ and $S=5.14$.  Using equation~\ref{eq-prec},
we can estimate the precession period assuming that $a_{22}$ 
time-averages to zero for sufficiently large $\Omega_p$.  We find a value
of $\dot{\phi}_{prec} = -0.14~\kmskpc$ which is comparable in
magnitude to our chosen pattern speeds but opposite in sign. This halo
corresponds to our choice of $q_{tidal}$ of about 6.3. We then
integrate the Euler equations for approximately 53 Gyr corresponding
to a little more than a typical precession period to get a complete
picture of the evolution of the disk for comparison to the N-body
simulations. Figure~\ref{fig-xi-vs-eta} shows the evolution of the
spin axis of the disk for comparison to the similar plots for the
N-body simulation (Fig.~\ref{fig-inc}). As in Fig.~\ref{fig-inc},
Figure~\ref{fig-xi-vs-eta} presents this evolution by tracing the
normalized, $x$ component of the disk angular momentum against the $y$
component of the same.

The agreement in behavior for the four pattern speeds is excellent
suggesting that the N-body disk is behaving in a similar manner to a
rigid body over this time period.  The nutation induced by the
triaxiality is readily apparent again in this plot. We notice that the
nutation period is shorter for a higher pattern speed. Thus, for
larger pattern speeds, as the nutation period is small, we might
expect a spiral galaxy to wobble several times over its lifetime if it
is embedded in a rapidly tumbling dark halo.  For the slower pattern
speeds, the nutation period becomes comparable to the precession
period and the resulting effect is simply a steady change of the disk
orientation.  Stars and gas on the outer edge of the disk that have
weaker self-gravity are prone to precess differentially and this may
be an additional origin of galactic disk warps.

\subsection{Resonant Interactions}
For the flattened halo discussed here the precession rate of disk is
negative while the tumbling rate is positive and in the same sense as
the disk.  While this case is the most reasonable, it is possible that
the sign of the precession rate and halo tumbling rate might be the
same and so resonances can occur when $\Omega_p = \dot{\phi}_{prec}$.
In the first investigations of orbits in a rotating 
triaxial potential, \citet{bin78, bin81} discovered that under
certain conditions retrograde orbits in the rotating frame
can be unstable leading to large $z$-motions 
due to resonant interactions with the
potential and suggested that this might be a mechanism to produce galactic
warps.  \citet{heisler_82} demonstrated the existence of 
Binney's resonance in orbital studies using an analytic triaxial potential.
\citet{tre00} have even suggested that this resonance may provide a mechanism
for creating polar ring galaxies.   
Accreted gas and forming stars counter-rotating with respect to main disk
may be levitated out of the plane if there is an underlying
rotating triaxial potential. 

So far the study in this paper has elucidated how self-consistent 
N-body disks evolve in tumbling triaxial potential rotating in 
the same direction as the disk.  In the other case of a counter-rotating
disk, it is interesting to see if Binney's resonance manifests itself in
disk evolution in our models.

Let us first consider the simple rigid disk model.
Consider the Euler equations again for the case where $\ddot{\phi}=0$
and $\dot{\theta} =0$ corresponding to steady precession at a fixed
angle $\theta$ at a constant rate.  For this to be the case in general,
we need to find the condition such that $\partial V/\partial\phi=0$ at
all times.  
Differentiating equation~\ref{eq-pot} with respect to
$\phi$ and setting it to zero implies that $\sin[2(\Omega_p t -
\phi)]=0$.  If $\phi = \dot{\phi}_{prec} t$ then this implies a
resonance when $\Omega_p = \dot{\phi}_{prec}$.  
With this resonant condition, we can use the first Euler equation to
find $\dot{\phi}$ from the quadratic equation
\begin{equation}
-I_1 \dot{\phi}^2 \sin\theta \cos\theta + S\dot{\phi}\sin\theta + \frac{\partial V}{\partial\theta} = 0
\end{equation}
where $\frac{\partial V}{\partial\theta}$ is known from
equation~\ref{eq-torque-appendix} (see Appendix~1).  For a
flat rotation curve with speed $v_c$ the disk angular momentum
is given by $S=2 M_d R_d v_c$.
If we assume that
$\dot{\phi}_{prec}$ is small then to a good approximation it is given by
\begin{equation}
\dot{\phi}_{prec} \approx -\frac{9}{2} \frac{R_d}{v_c} ( a_{20} + 2 a_{22}) \cos\theta.
\label{eq-prec-res}
\end{equation}
For our model of M31, with our estimate of $a_{20}$ and $a_{22}$ for
the coefficients of an external quadrupole potential and an initial
disk inclination of $\theta=30^\circ$, the resonance occurs when
$\Omega_{p,res} = \dot{\phi}_{prec} = -0.055~\kmskpc$. Thus we see
that the size of this typical resonant precession frequency is
comparable to the expected tumbling frequencies of dark halos.  

\begin{figure}
\plotone{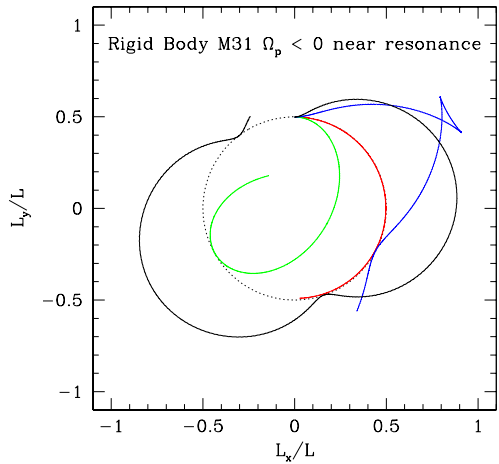}
\caption{{The trace of the spin axis measured by the
	normalized components of the disk angular momentum vector
    for the
    rigid body model of M31 forced by an external quadrupole potential
    near the resonant pattern speed.  Colored lines correspond to the
    pattern speeds $\Omega_p=\Omega_{p,res}$ (red);
    $\Omega_p=\Omega_{p,res}/2$ (green); $\Omega_p= 2\Omega_{p,res}$
    (blue); and $\Omega_p=4 \Omega_{p,res}$ (black). The resonance
    describes a region of unstable behavior where the disk can undergo
    large changes in inclination with irregular precession and
    nutation.}
\label{fig-resonant}
}
\end{figure}

We can examine the behavior of the disk precessional evolution near
this resonance.  In Fig.~\ref{fig-resonant}, we represent the disk
spin axis evolution to bring out the unstable behavior of rigid models
with $\Omega_p$ set to $\Omega_{p,res}/2, \Omega_{p,res},
2\Omega_{p,res}$ and $4\Omega_{p,res}$.  The radically different
near-resonance disk behavior noted in this figure, marks the position
of a region of instability. Here the disk inclination can change by
large amount and the precession and nutational behavior are quite
irregular.

\begin{figure}
\plotone{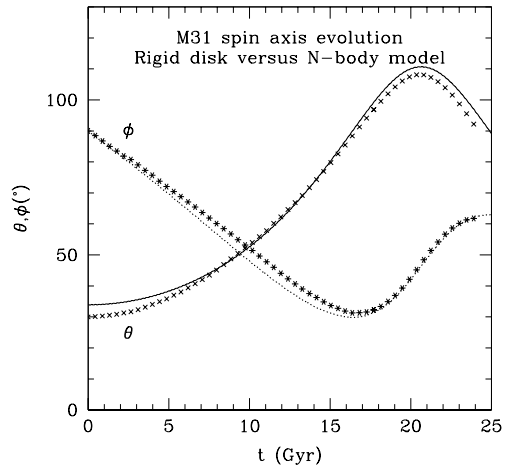}
\caption{The time evolution of the polar angles $(\theta,\phi)$ of the
direction of the disk angular momentum vector for an M~31 rigid disk model
(smooth curves)
and N-body model (points).  In both cases, the external quadrupole potential is
counter-rotating with $\Omega_p = -0.11~\kmskpc$, a value that is twice the
resonant frequency for this model.  The agreement between the rigid disk
model and N-body simulation is excellent when we adjust the initial
inclination to $\theta=34^\circ$ for the rigid disk case.  In this case,
the disk inclination increases by almost 90$^\circ$ before reversing
direction.
\label{fig-theta-phi}
}
\end{figure}

We also consider an additional M31 N-body model for comparison to the 
rigid disk model.  We evolve it in the same way as previous models but here
examine a counter-rotating external quadrupole potential 
with $\Omega_p=-0.11~\kmskpc$ to see if the irregular behavior that is
seen in the simple rigid disk model is reproduced in a self-consistent  
N-body model.   Figure~\ref{fig-theta-phi} shows a comparison of 
the time evolution of the polar angles $(\theta,\phi)$ representing 
the direction of the disk angular momentum vector 
for the rigid disk and
N-body disk models. (Here $\phi$ is offset by 90 degrees from the usual
definition as the angle of the line of nodes in the Euler equation.) 
For the rigid disk, we adjust the initial inclination
slightly to $\theta=34^\circ$ and find excellent agreement with 
the N-body disk evolution.   We have integrated for the unusually long time
of 25 Gyr to see the full tilt evolution.  However, 
the amplitude of the quadrupole potential could easily be twice as large
based on the observed variance of $q_{tidal}$ in cosmological halos and so
this behavior would occur in half the time since from
equation~\ref{eq-prec-res} we see that the precession rate is directly
proportional to the external quadrupole amplitude.
It is remarkable that in this counter-rotating case, 
the disk inclination grows by 90$^\circ$ before reversing course.

While the Binney resonance is
derived in the context of closed collisionless orbits in rotating
triaxial potentials, a version of it also appears to be operating 
collectively in a self-gravitating disk in the same context.
Our idealized model of a N-body disk forced by a rotating external 
quadrupole potential tips through 90$^\circ$ when counter-rotating 
tumbling frequencies are near the resonant value.   The case of
counter-rotation is probably rare but not impossible given the existence of
some disks with counter-rotating components \citep[e.g.,][]{rub92}.  It may
also help explain the phenomenon of polar rings as suggested by
\citet{tre00}.   Accreted gas that is counter-rotating with respect to the
tumbling halo might rise out of the halo principal plane within a Hubble
time given a large enough quadrupole amplitude and suitable tumbling
frequency.  One caveat to be aware of is that given the long timescales 
of this process, dynamical friction may begin to align even the outer 
parts of the halo so the resulting torque may diminish with time and 
so weaken the effect.

\section{Conclusions}
\noindent
In this paper we discuss the effect of the tidal field of a tumbling
and triaxial external halo on the dynamical evolution of a galactic
disk. We have shown that the disk precesses in response to its
misalignment with the external halo and the gradual re-alignment of
the disk with the external tumbling halo causes it to warp. The disk
is also seen to develop strong spirality to its edge in the potential
of the external halo that it sits in.  It is also suggested that if a
disk feels a significant external quadrupole that is essentially
static or at most slowly tumbling, it might trigger a bar instability,
late in the disk. 

The conclusions achieved in this work are based on analytical
calculations that are backed up by N-body simulations of the Milky Way
and M~31 models, subject to the quadrupolar tug of an external
tumbling and triaxial halo. The quadrupolar strength is judged via
cosmological simulations, while a range of halo tumbling periods are
scanned through in the simulations. We advance this dynamical
mechanism as an important cause of non-axisymmetric structures to set
in galactic disks. An important result of our work is that the
prodding of the disk by a external quadrupole may not always produce
bars in the disks. Thus, the M~31 disk was found to be robust to the
bar instability, though the Milky Way disk was found to be much more
susceptible to the bar being triggered by this mechanism. However,
both disks were found to warp. The dispersion in cosmological halo
properties imply that the external quadrupole is rather weak in some
systems and so may not be effective in driving disk evolution.

Our work elucidates the importance of the quadrupolar term in the
potential of the external halo, particularly, in terms of the
trigerring of the bar and warp instabilities. Given the viability of
such tidal prodding of the disk by the external halo, detailed
investigation of this mechanism in galaxy formation simulations
is suggested.

\section*{Acknowledgments}
We acknowledge help from the CITA visitor programme which aided in this
collaboration. We thank an anonymous referee for detailed comments that
greatly improved this paper.  We acknowledge NSERC for financial assistance 
and SHARCNET for providing supercomputer time. DC acknowledges the support 
of a Royal Society Dorothy Hodgkin Research Fellowship.

\appendix

\section[]
\noindent

We calculate the interaction potential between an exponential disk and a
time-dependent external quadrupole potential rotating with pattern speed
$\Omega_p$.  If we first assume the potential is static and aligned with 
the body axes $(x',y',z')$ then  we can write the potential from
equation~\ref{eq-phi-ext} in Cartesian
coordinates as:
\begin{equation}
\Phi_{ext}(x',y',z') = b_1 x'^2 + b_2 y'^2 + b_3 z'^2
\end{equation}
with
\begin{eqnarray}
x'&=& r \cos\phi \sin\theta \\
y'&=& r \sin\phi \sin\theta \\
z'&=& r \cos\theta
\end{eqnarray}
where one can show easily that 
the coefficients $b_1$, $b_2$ and $b_3$ are defined in terms of the
harmonic expansion in equation~\ref{eq-phi-ext} through:
\begin{eqnarray}
b_1 &=& -a_{20}/2 + 3a_{22} \\
b_2 &=& -a_{20}/2 - 3a_{22} \\
b_3 &=& a_{20}
\end{eqnarray}
If the potential is rotating counterclockwise about the $z'$-axis with pattern speed
$\Omega_p$
then we can introduce the time-dependence of the potential into the
coordinates $(x',y',z')$ using inertial frame coordinates $(x,y,z)$ where
\begin{eqnarray}
x' &=& x \cos \Omega_p t - y \sin \Omega_p t \\
y' &=& x \sin \Omega_p t + y \cos \Omega_p t \\
z' &=& z
\end{eqnarray}
so that potential becomes
\begin{equation}
\Phi_{ext}(x,y,z; t) = d_1(t) x^2 + d_2(t) y^2 + d_3(t) xy + d_4 z^2
\end{equation}
with the coefficients
\begin{eqnarray}
d_1(t) &=& -a_{20}/2 + 3a_{22} \cos 2\Omega_p t\\
d_2(t) &=& -a_{20}/2 - 3a_{22} \cos 2\Omega_p t\\
d_3(t) &=& 6 a_{22} \sin 2\Omega_p t \\
d_4 &=& a_{20}
\end{eqnarray}

We can thus derive the time-dependent interaction potential between 
a tilted exponential disk and this tumbling quadrupole potential.
The parametric equation of a ring of radius $R$ with inclination angle $\theta$ and line
of nodes $\phi$ is given by:
\begin{eqnarray}
x &=& R (\cos\phi\cos\psi - \sin\phi\cos\theta\sin\psi) \\
y &=& R (\sin\phi\cos\psi + \cos\phi\cos\theta\sin\psi) \\
z &=& R \sin\theta\sin\psi
\label{ring:parametric}
\end{eqnarray}
where $\psi$ is the angle measured along the ring.
The mass element of the ring is given by
\begin{equation}
dM = \Sigma(R) R\; dR\; d\psi
\end{equation}
so we can derive the interaction potential by integrating over $\psi$ and
$R$ through
\begin{equation}
V(\theta, \phi; t) = \int d\psi \int \Sigma(R) R\; dR \; \Phi_{ext}(x,y,z)
\end{equation}
If we assume an exponential disk with
\begin{equation}
\Sigma(R) = \frac{M_d}{2\pi R_d^2} e^{-R/R_d}
\end{equation}
and substitute the expressions for $(x,y,z)$ in equations
\ref{ring:parametric} and do the integrals we arrive at the 
intermediate form of the interaction potential:

\begin{equation}
V(\theta,\phi; t) = 3 M_d R_d^2 [ d_1(t)(1 - \sin^2 \phi \sin^2 \theta) + d_2(t) (1
- \cos^2 \phi \sin^2 \theta) + 
+ d_3(t) (\sin 2\phi \sin^2 \theta)/2
+ d_4 \sin^2 \theta ]
\end{equation}

This can be reduced to the following simpler form in terms of the original
quadrupole coefficients
$a_{20}$ and $a_{22}$:
\begin{equation}
V(\theta,\phi; t) = 3 M_d R_d^2 \left \{ -\frac{a_{20}}{2}(3\cos^2\theta - 1) +
3a_{22}\cos[2(\Omega_p t - \phi)] \sin^2\theta \right \}
\end{equation}

The derivatives of the interaction potential used in the Euler equations are then:

\begin{eqnarray}
\frac{\partial V}{\partial \theta} &=& 9 M_d R_d^2 \sin\theta\cos\theta \left \{ a_{20} + 2
a_{22} \cos[2(\Omega_p t - \phi)] \right \} \\
\label{eq-torque-appendix}
\frac{\partial V}{\partial \phi} &=& 18 M_d R_d^2 a_{22}\sin[2(\Omega_p t - \phi)] \sin^2\theta
\end{eqnarray}

\vspace{1cm}
\noindent
\bibliographystyle{apj}


\end{document}